\documentclass[aps,pre,twocolumn,superscriptaddress]{revtex4}
\usepackage{graphicx}
\usepackage{color}
\usepackage{amsmath}

\begin{document}

\title{Resetting processes with non-instantaneous return}
\author{Anna S. Bodrova}
\affiliation{Humboldt University, Department of Physics, Newtonstrasse 15, 12489 Berlin, Germany}
\author{Igor M. Sokolov}
\affiliation{Humboldt University, Department of Physics, Newtonstrasse 15, 12489 Berlin, Germany}
\affiliation{IRIS Adlershof, Zum Gro{\ss}en Windkanal 6, 12489 Berlin, Germany}

\begin{abstract}
We consider a random two-phase process which we call a reset-return one. The particle starts its motion at the origin. 
The first, displacement, phase corresponds to a stochastic motion of a particle and is finished at a resetting event. 
The second, return, phase corresponds to the particle's motion towards the origin from the position it 
attained at the end of the displacement phase. This motion
towards the origin takes place according to a given equation of motion. The whole process is a renewal one. 
We provide general expressions for the stationary probability density function of the particle's position and for the mean hitting 
time in one dimension. We perform explicit analysis for the Brownian motion during the displacement phase and three different types of the return motion: return at a constant speed, return at a constant acceleration with zero initial speed and return under the action of a harmonic force. We assume that the waiting times for resetting events follow an exponential distribution, or that resetting takes place at a constant pace. For the first two types of return motion and the exponential resetting the stationary probability density function of the particle's position is invariant under return speed (acceleration), while no such invariance is found for deterministic resetting, and for exponential resetting with return under the action of the harmonic force. We discuss necessary conditions for such invariance of the stationary PDF of the positions with respect to the properties of the return process, and demonstrate some additional examples when this invariance does or does not take place.
\end{abstract}

\maketitle

\section{Introduction} 
Random motion under resetting \cite{review} represents a process in which a stochastically moving particle returns from time to time to its 
initial position and starts its motion anew from the very beginning. Examples of such processes are found in many fields such as
biochemistry \cite{chemistry,bio1,bio2}, biology \cite{biology,bio3} and computer science \cite{computerscience}.
In computer science, random walks with stochastic restarts represent a useful strategy to optimize search
algorithms in computationally  hard problems \cite{computerscience}. The organisms can use stochastic resetting
or switching between different phenotype states in order to adapt to fluctuating environments \cite{biology}.
Also the motion of foraging animals, when they are searching for food and return home from time to
time is one of the example of resetting. One can also consider a robotic vacuum cleaner performing stochastic
motion while cleaning the apartment and returning to its base in order to recharge.

Early works concentrated on the case when the stochastic displacement process is a simple Brownian motion
\cite{EvansMajumdar}, i.e. a Markovian process with stationary increments. Later on, one has also studied other types of motion
between the resetting events, such as L\'evy flights \cite{levy1,levy2}, continuous-time random walks with or
without drift \cite{MV2013,MC2016,Sh2017,ctrw}, and scaled Brownian motion \cite{Anna01, Anna02}.
The waiting times between the resetting events are considered to be distributed according to an exponential distribution \cite{EvansMajumdar}, 
$\delta$-distribution (resetting at a fixed time after starting the stochastic motion) \cite{shlomi2017, palrt}, power-law \cite{NagarGupta} and other types  \cite{res2016,shlomi2016}.

The main attention was always paid to the probability density function (PDF) of displacements as measured 
at a given time or in a stationary state, and to the mean first passage time (MFPT) --
the average time necessary to hit a specified target \cite{redner,redner2}. With respect to the first hitting properties, early work concentrated on searchers performing one-dimensional 
Brownian motion with Poissonian resetting \cite{EvansMajumdar,EM2011,evma13}. While the
MFPT to a target for a diffusing particle in absence of resetting may diverge, in presence of resetting it turns
finite, and there exists an optimal rate of resetting which minimizes
the MFPT. The discussion has been extended to two and higher dimensions in \cite{high12,bhat}. It has been shown that 
the search process is most effective in the case of the deterministic resetting, occurring with constant time intervals between the resetting events \cite{SokChech, shlomi2017, palrt}.

\begin{figure}[ht]
\includegraphics[width=0.98\columnwidth]{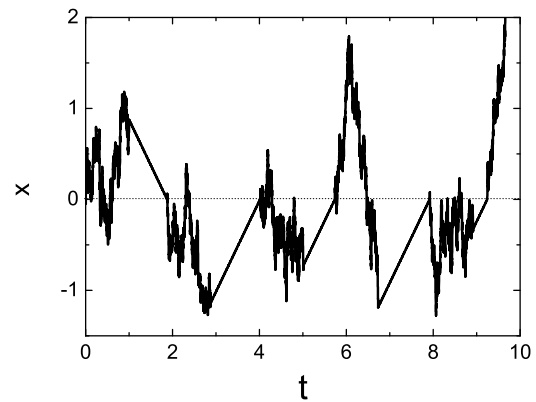}
\caption{A trajectory of a particle performing Brownian motion under resetting with the return at a constant  speed $v=1$. The resetting events follow at a 
fixed time $t_r=1$ after the beginning of the displacement phase. The diffusion coefficient in the displacement phase is $D=1$.} \label{fig:trajectories} 
\end{figure}

Typically, the return to the initial position (assumed at the origin of the coordinate system in what follows) 
is considered to be instantaneous. In some studies, a random refractory period after the resetting event has 
been introduced \cite{wait1,wait2,wait3}, but the return to the origin was still described as an instantaneous jump. 
However, in many situations, especially when the process considered corresponds to a motion of a material object in space, 
such an assumption is nonphysical. In what follows we discuss what happens when the return to the origin follows
a given equation of motion, and takes finite time for completion. The exemplary trajectories of such a process are depicted at Fig.~\ref{fig:trajectories}
for the case of Brownian motion with return at a constant speed in one dimension (1d). This situation was just recently 
considered by Arnab Pal, {\L}ukasz Ku\'smierz and Shlomi Reuveni \cite{shlomi,shlomi1,shlomi2}, which work was performed parallel to ours, and the comparison to their work is given below. 

In the current study we derive the stationary PDF and the mean squared displacement (MSD) as well as the 
MFPT for the non-instantaneous return to the origin.  We proceed as follows. In the next Section II we define the model of our system 
and give general analytic expressions for the stationary PDF. 
In Sections III and IV we consider the normal diffusion under exponential and deterministic resetting, correspondingly. In Section V we discuss the conditions under which the PDF 
remains invariant with respect to the return speed and consider the examples of situations when such an invariance takes place for any resetting time distribution,
or when no such resetting time distribution exists. Section VI is devoted to the investigation of the MFPT. Finally, we give our conclusions in Section VII.

\section{Probability density function} 
\subsection{Model}
\begin{figure}[ht] 
\includegraphics[width=0.85\columnwidth]{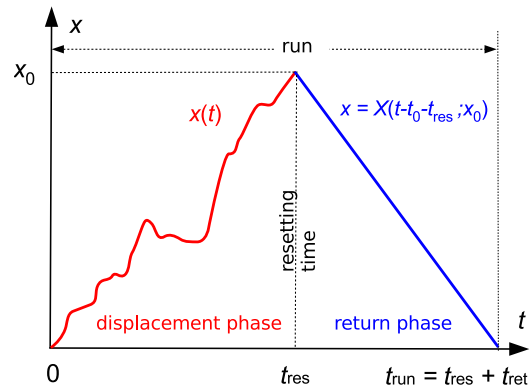}
 \caption{The notation used. A run stars at time $t_0$ 
with a renewal event and consists of two motion phases: the displacement phase and the return phase. The time
axis corresponds to the time elapsed from the beginning of the run $t-t_0$.
The total duration of the return phase $t_{\rm ret} (x_0)$ depends on the particle's position $x_0$ 
at the instant of resetting and on the equation of motion $x=X(t;x_0)$. \label{Gepoch}} 
\end{figure}

The reset-return process consists of subsequent runs. Each run is a sequence of two processes, the
stochastic displacement process $x(t)$ which is interrupted by the resetting event, and the deterministic return
process, which ends when the particle returns to the origin, as depicted in Fig.~\ref{Gepoch}. The waiting time 
density for a resetting event is given by a function $\psi(t_{\mathrm{res}})$, where $t_{\mathrm{res}}$ is the 
time elapsed from the beginning of the run. The motion during the return phase takes place according to the 
deterministic equation of motion $x = X(t;x_0)$, where $t$ now is the time elapsed since the beginning of the 
return phase, and $x_0$ is the particle's coordinate at the beginning of this phase (i.e. at the end of the 
preceding displacement one). The equation of motion $X(t;x_0)$ is chosen such, that the return from
each point $x_0$ to the origin takes place in a finite time $t_{\rm ret} (x_0)$, given by the solution of the equation
\begin{equation}\label{Xtret}
X(t_{\rm ret} (x_0);x_0) = 0\,. 
\end{equation}
The full run finishes by the return to the origin. Then the process restarts 
anew with the displacement phase of the next run. Therefore the whole process is a renewal one. The total 
duration of the run is $t_{\mathrm{run}} = t_{\mathrm{res}} + t_{\mathrm{ret}}$. 

\subsection{Duration of the run}
Let $\phi(t)$ denote the probability density that the duration of the run is equal to $t$. It is convenient to work in the Laplace domain. The Laplace transform of the resetting PDF is
\begin{equation}
\tilde{\phi}(s)=\int_0^{\infty}\phi(t)\exp(-ts)dt\,.
\end{equation}
The probability density $\phi_n(t)$ for the particle to return to the origin during the $n$-th run at time $t$ satisfies the renewal equation \cite{sokbook}
\begin{equation}
\phi_n(t) = \int_0^{t}\phi_{n-1}(t^{\prime})\phi(t-t^{\prime})dt^{\prime}\,.
\end{equation}
Here we assume that the previous run has ended at time $t^{\prime}$ and the duration of the very last run is $t-t^{\prime}$. The sum of all $\phi_n(t)$ gives the renewal rate $\kappa(t)$. 
\begin{equation}
\kappa(t)=\sum_{n=1}^{\infty}\phi_n(t)\,,
\end{equation}
which defines that any run has been terminated at time $t$. The Laplace transform of $\kappa(t)$ yields
\begin{equation}
\tilde \kappa (s) = \sum\limits_{n = 1}^\infty  {{{\tilde \phi }_n}(s)}
  = \frac{{\tilde \phi (s)}}{{1 - \tilde \phi (s)}}\,.
\label{phis}
\end{equation}

In the present work we will concentrate only on the situation when the first moment $\langle t_{\rm run} \rangle = \int_0^\infty t \phi(t) dt$ of $\phi(t)$  does exist and its Laplace transform is given by 
\begin{equation}
\langle \tilde{t}_{\rm run} \rangle = -\lim_{s \to 0}\frac{d}{ds}\int_0^{\infty}\phi(t)\exp(-ts)dt=-\frac{d\tilde{\phi} (s)}{ds} \Bigr|_{s=0} .
\end{equation}
In such a way we get $\tilde{\phi} (s)=1-\langle t_{\rm run} \rangle s+\ldots$. Using Eq.~(\ref{phis}) and performing the inverse Laplace transform, we obtain 
\begin{equation}
\label{kappat}
\lim_{t \to \infty}\kappa(t)=\frac{1}{\langle t_{\rm run} \rangle} .
\end{equation}
If the first moment $\langle t_{\rm run} \rangle$ doesn't exist, $\kappa(t)$ stays explicitly time-dependent. 
Such general reset-return processes seem to be rich with respect to possible aging phenomena, whose investigation is left for further work. 

The distribution $\psi(t_{\mathrm{res}})$ of  resetting times $t_{\mathrm{res}}$ is given, and the return time $t_{\rm ret} (x_0)$ 
is the deterministic function of the coordinate $x_0$ at the end of the displacement phase. Therefore
\begin{eqnarray}\nonumber
&& \phi(t_{\mathrm{run}}) = \int_0^\infty d t_{\mathrm{res}} \psi(t_{\mathrm{res}}) \times \\
&&\times \int_{-\infty}^\infty d x_0 \delta[t_{\mathrm{run}} - t_{\mathrm{res}} - t_{\mathrm{ret}}(x_0)] p(x_0|t_{\mathrm{res}}).\label{phirun}
\end{eqnarray}
Here $p(x|t)$ is the PDF of the displacement during the displacement phase at time $t$.

The mean value of the run duration $\langle t_{\mathrm{run}}\rangle$ is given by the sum of the duration of the resetting $\langle t_{\mathrm{res}} \rangle$ and return $\langle t_{\mathrm{ret}} \rangle$ phases 
\begin{eqnarray}\label{trun}
 &&\!\!\!\!\!\langle t_{\mathrm{run}} \rangle = \langle t_{\mathrm{res}} \rangle + \langle t_{\mathrm{ret}} \rangle ,\\ \label{tres}
 &&\!\!\!\!\!\langle t_{\mathrm{res}} \rangle= \int_0^\infty t' \psi(t') dt' , \\
&&\!\!\!\!\!\langle t_{\mathrm{ret}} \rangle=\int_0^\infty dt' \int_{-\infty}^\infty d x_0 t_{\mathrm{ret}}(x_0) p(x_0|t') \psi(t'),\label{tret}
\end{eqnarray}
where $t_{\mathrm{ret}}$ is given by the solution of Eq.~(\ref{Xtret}).

\subsection{General form of the probability density function}

Our next task will be to obtain the displacement PDF at time $t$
\begin{equation}
 P(x,t) = \int_0^t \kappa(t_0) Q(x; t - t_0) dt_0\,,
 \label{eq:Poft}
\end{equation}
 where $Q(x;t)$ is the PDF of the position at the measurement time $t$, and $\kappa(t_0)$ accounts for the fact that the last run has started at time $t_0$. 
 Let us first take the beginning of the displacement phase at $t_0$ as a new origin of time. The measurement time is now $\Delta t = t - t_0$. 
 For given $t_{\mathrm{res}}$ and $x_0$ the cases when this measurement time $\Delta t$ falls into the displacement or into the return phases are mutually excluding. Let us first 
fix the value of $t_{\mathrm{res}}$ (i.e. consider all realizations with reset time in the close vicinity of $t_{\mathrm{res}}$) 
and get the PDF of displacements conditioned on $t_{\mathrm{res}}$ and on the corresponding $x_0$:
\begin{eqnarray}\label{qqq}
q(x|\Delta t;t_{\mathrm{res}},x_0) &=& q_1(x|\Delta t;t_{\mathrm{res}})+q_2(x|\Delta t;t_{\mathrm{res}},x_0)\,, \nonumber \\
q_1(x|\Delta t;t_{\mathrm{res}}) &=& p(x|\Delta t) \Theta(t_{\mathrm{res}} - \Delta t)\,, \\
q_2(x|\Delta t;t_{\mathrm{res}},x_0) &=&\delta[x - X(\Delta t-t_{\mathrm{res}};x_0)] \times  \nonumber \\
&& \Theta(\Delta t-t_{\mathrm{res}} ) \Theta[t_{\mathrm{res}} + t_{\rm ret} - \Delta t]\,. \nonumber
\end{eqnarray}
Here the $\Theta$-function in $q_1$ represents the condition that the measurement time falls into the displacement phase of the run. The first $\Theta$-function in 
$q_2$ accounts for the measurement performed during the return phase and the last $\Theta$-function in $q_2$ gives the condition that the measurement occurs before 
the particle returns to the origin. To lift the conditioning on $x_0$ and $t_{\mathrm{res}}$ we first average over the distribution of $x_0$ for given $t_{\mathrm{res}}$ 
and then over the distribution 
of $t_{\mathrm{res}}$:
\begin{eqnarray}\label{Q1Q2}
Q(x;\Delta t) &=& Q_1(x;\Delta t)+Q_2(x;\Delta t)\,,\\
Q_1(x;\Delta t) &=& \int_0^\infty  dt_{\mathrm{res}}\psi(t_{\rm res}) q_1(x|\Delta t;t_{\mathrm{res}})\,,  \nonumber \\
Q_2(x;\Delta t) &=& \int_0^\infty dt_{\mathrm{res}}\psi(t_{\rm res}) \times \nonumber \\
 && \int_{-\infty}^{\infty}dx_0 q_2(x|\Delta t;t_{\mathrm{res}},x_0)p(x_0|\Delta t_{\mathrm{res}})\,. \nonumber
\end{eqnarray}
The normalization of $P(x,t)$ given by Eq.~(\ref{eq:Poft}) is proved in the Appendix \ref{Anorm}. The MSD can be calculated performing the integration:
\begin{equation}\label{msdpdf}
\left\langle x^2\right\rangle = 2\int_0^{\infty} x'^2 P(x^{\prime})dx^{\prime}\,.
\end{equation}

\subsection{Stationary probability density function}
In the present work we concentrate on stationary PDF in the asymptotic limit $t \to \infty$
\color{black}
\begin{equation}
 P(x) = \lim_{t \to \infty} P(x,t)= \int_0^{\infty} \kappa(t_0) Q(x; t - t_0) dt_0.
\end{equation} 
The consideration of non-stationary PDFs is beyond of the scope of the current paper. 

Although one cannot exclude that a stationary PDF of coordinate does exist also for non-stationary process of renewals, 
the situation is really simple only if the renewal resetting process does possess the mean waiting time and is stationary. 
In this case $\kappa(t_0)$ is given by Eq.~(\ref{kappat}) for $t_0$ sufficiently long, and the expression for $P(x)$ reads
\begin{equation}\label{mixture}
 P(x) = \frac{\rho_1(x) + \rho_2(x)}{\langle t_{\rm res}\rangle+\langle t_{\rm ret} \rangle}
\end{equation}
with $\rho_1(x) = \int_0^\infty Q_1(x;t') dt'$ and $\rho_2(x) = \int_0^\infty Q_2(x;t') dt'$. 
Eq.~(\ref{mixture}) represents a mixture of probability density functions of positions in corresponding phases of motion taken with weights proportional to the duration of the phases. $P_1(x)=\rho_1(x)/\langle t_{\rm run}\rangle$ and $P_2(x)=\rho_2(x)/\langle t_{\rm run}\rangle$ account for the probability densities governing the displacement and return phases, correspondingly. Thus, $\rho_1(x)$ and $\rho_2(x)$ are the rescaled conditional PDFs of the displacement and return phase, correspondingly (PDF divided by $\langle t_{\rm run}\rangle$).

The mean duration of the resetting phase is
\begin{equation} \label{eq:IntI1}
\langle t_{\rm res}\rangle=\int_0^{\infty}\rho_1(x)dx\,,
\end{equation}
and the mean duration of the return phase is
\begin{equation}\label{tretI}
\langle t_{\mathrm{ret}} \rangle=\int_0^{\infty}\rho_2(x)dx\,.
\end{equation}
If the fraction
\begin{equation}\label{I1I2C}
\frac{\rho_2(x)}{\rho_1(x)}=C\,,
\end{equation}
where $C=\rm const$, then the PDF becomes independent of the return process:
\begin{equation}
 P(x) = \frac{\rho_1(x)}{\langle t_{\rm res} \rangle} \,.
 \end{equation}
Below we will investigate under which conditions does this statement hold.

\subsection{PDF of the displacement phase}

The rescaled PDF of the displacement phase $\rho_1(x)$ is given by
\begin{equation}\label{I1}
 \rho_1(x)=\int_0^\infty dt_{\mathrm{res}} \psi(t_{\mathrm{res}}) \int_0^{t_{\mathrm{res}}} p(x|t) dt 
\end{equation}
and can be immediately evaluated since $\psi(t)$ and $p(x|t)$ are known. In the following we consider a  displacement process corresponding to standard Brownian motion (pure diffusion without drift) with a PDF 
\begin{equation}\label{pgau}
p(x|t)=\frac{1}{\sqrt{4 \pi D t}} e^{- \frac{x^2}{4 D t}}
\end{equation}
We also investigate two different types of the waiting time distributions of the resetting events: exponential waiting time distribution: $\psi(t)=re^{-rt}$ and deterministic resetting with fixed time interval: $\psi(t)=\delta(t-t_r)$.

\subsection{PDF of the return phase}

For the rescaled PDF of the return phase $\rho_2(x)$ we obtain
\begin{eqnarray}
  \rho_2(x)= && \int_0^\infty\! dt' \int_0^\infty \! d t_{\mathrm{res}} \psi(t_{\mathrm{res}}) \times \nonumber \\
 && \int_{-\infty}^\infty\! dx_0 \delta[x - X(t'-t_{\rm res};x_0)] \times  \label{I2} \\
&& \Theta(t'-t_{\mathrm{res}}) \Theta[t_{\mathrm{res}} + t_{\rm ret} (x_0)-t'] p(x_0|t_{\mathrm{res}}).  \nonumber
\end{eqnarray}
We assume that the equation of motion is such, that the distance to the origin in the return phase is a monotonically decreasing
function of time. We moreover assume that the equation of motion $x=X(t-t_{\rm res};x_0)$ is symmetric with respect to $x_0$: For $x_0 > 0$ the value of $x$ stays non-negative all the time
until it reaches the value $x=0$; for $x_0 < 0$ it stays negative all the time, and $X(t-t_{\rm res};- x_0) = - X(t-t_{\rm res}; x_0)$.

Then $\rho_2(x)$ can be presented as sum of $\rho_2^{+}(x)$ for $x>0$ and $\rho_2^{-}(x)$ for $x<0$. 
In the following we will discuss the evaluation of $\rho_2^+(x)$, while the evaluation of $\rho_2^-$ follows along the similar lines.
Thus, we assume  $x>0$ and interchange the sequence of integration in $t'$ and in $t_{\mathrm{res}}$ in Eq.~(\ref{I2}). We note that the first $\Theta$-function does not depend on $x_0$, can be safely moved out from the last integral,
and then essentially fixes the lower bound of integration in $t'$. Then we introduce the new variable of integration $\tau = t' - t_{\mathrm{res}}$ and get
\begin{eqnarray}
 \rho_2^+(x) = \int_0^\infty dt_{\mathrm{res}} \psi(t_{\mathrm{res}} )\int_0^\infty dx_0   p(x_0 | t_{\mathrm{res}})\times\\\nonumber \times\int_0^\infty d\tau \delta[x - X(\tau;x_0)]\Theta(t_{\mathrm{ret}}(x_0)-\tau).
\end{eqnarray}
Now we note that $t_{\mathrm{ret}}(x_0)$ is essentially the solution of the equation $X(\tau;x_0) = 0$ for given $x_0$. Thus, the last $\Theta$-function is always unity when the argument of the $\delta$-function is zero, and therefore can be omitted:
\begin{eqnarray}
 \rho_2^+(x) &=& \int_0^\infty dt_{\mathrm{res}} \psi(t_{\mathrm{res}} ) \times \label{Iplus} \\
 && \int_0^\infty dx_0   p(x_0 | t_{\mathrm{res}}) \int_0^\infty d\tau \delta[x - X(\tau;x_0)].  \nonumber
\end{eqnarray}
From our monotonicity assumption it follows that the function $X(t;x_0)$ is a monotonically decaying function of time $t$ (the particle steadily approaches the origin) and a monotonically growing
function of $x_0$ (the further from the origin the initial position is, the longer the particle needs to return to the origin). This means that the 
equation $x - X(t'-t_{\mathrm{res}};x_0) = 0$ does not possess roots for $x > x_0$ and does possess a single root for $0 < x < x_0$. Therefore the integral given by Eq.~(\ref{Iplus}) is zero for $x > x_0$. 
Let now   $\tau(x,x_0)$ be the solution of the equation $X(\tau;x_0) = x$ for $x < x_0$. This solution is given by the function inverse 
to $X(\tau;x_0)$. Now we use  the rule of variable change in the $\delta$-function and denote the speed at which the 
particle crosses $x$ on return from $x_0$ by $|v(x;x_0)|$:
\begin{eqnarray}
 \int_0^\infty d\tau \delta[x - X(\tau;x_0)] = \left\{
 \begin{array}{lll} 
  0 & \mbox{for} & x_0 < x \\
  |v(x;x_0)|^{-1} & \mbox{for} & x_0 > x .
 \end{array}
 \right.
\end{eqnarray}
The substitution of this expression into Eq.~(\ref{Iplus}) gives us the expression for $ \rho_2^+(x)$. Changing the sequence of integration, we obtain the final form for the rescaled PDF of the returm motion $ \rho_2(x)$
\begin{equation}\label{I2last}
 \rho_2(x)=\int_{\left|x\right|}^\infty \frac{d x_0}{\left|v(x;x_0)\right|} \int_0^\infty dt_{\mathrm{res}} \psi(t_{\mathrm{res}} )p(x_0 | t_{\mathrm{res}}) .
\end{equation}
Below we consider explicitly three different type of return motion.

\begin{figure}[ht]
\includegraphics[width=0.98\columnwidth]{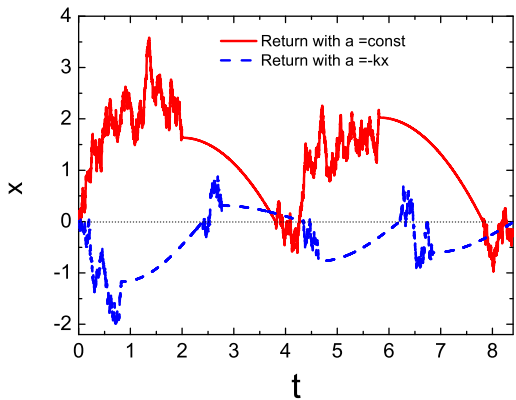}
\caption{Typical trajectories for the return with a constant acceleration $a=1$ (deterministic resetting with $t_r=2$) and acceleration, proportional to the displacement $a=-kx$, $k=1$ (exponential resetting with $r=1$).} \label{Gxyza}
\end{figure}

\subsection {Return at a constant speed $v=\rm const$} 
As a first example we consider the return at constant speed with
\begin{equation}\label{xvret}
x=x_0-vt .
\end{equation}
The duration of the return phase is linearly proportional to the absolute value of the starting position of the return motion 
\begin{equation}\label{ttretv}
t_{\rm ret}=\frac{\left|x_0\right|}{v}. 
\end{equation}
 Eq.~(\ref{I2last}) then attains the form
\begin{equation}  \label{I2vv}
\rho_2(x)=\frac{1}{v}\int_{|x|}^\infty dx_0 \int_0^\infty dt\; p(x_0|t) \psi(t) .
\end{equation}

\subsection{Return at a constant acceleration $a=\rm const$} 

We consider the return motion of a particle of unit mass $m=1$ under the action of a constant force or in a linear potential $U=ax$. We assume that at the beginning of the return motion the initial speed $v_0 = 0$ and constant acceleration $a > 0$ for $x_0>0$ both showing in the direction of the origin (the return motion for $x_0<0$ corresponds to $a < 0$):
\begin{eqnarray}\label{xxxret}
\left\{
 \begin{array}{l l}
 x(t) = x_0 -a t^2 / 2 \\
 v(t) = - a t .
  \end{array}
 \right. 
\end{eqnarray}
Here $t$ is time elapsed from the start of the return motion. The return time is  
\begin{equation}\label{ttreta}
t_{\rm ret}= \sqrt{\frac{2x_0}{a}}
\end{equation}
and the absolute value of the return speed is
\begin{equation}
 |v(x;x_0)| = \sqrt{ 2a(x_0 - x)}.
\end{equation}
 Eq.~(\ref{I2last}), in this case takes the form
\begin{equation}\label{I2a}
\rho_2(x)=\sqrt{\frac{1}{2|a|}}\int_{|x|}^{\infty}\frac{dx_0}{\sqrt{|x_0-x|}}\int_{0}^{\infty}dt_{\rm res}\psi(t_{\rm res})p(x_0|t_{\rm res})\,.
\end{equation}
The corresponding trajectories are shown at Fig.~\ref{Gxyza} with a red line.  

\subsection{Return under the action of a harmonic force}

Here we consider motion with acceleration proportional to the position of the particle: $a=-kx$ and zero initial velocity. We assume that a particle has a unit mass $m=1$ and is under the action of the harmonic spring with spring constant $k$ or in a quadratic potential $U=kx^2/2$. The coordinate during the return process is 
\begin{equation}\label{xharm}
x=x_0 \cos\left(\sqrt{k}t\right)\,, 
\end{equation}
where $t$ is time elapsed from the start of the return motion. The absolute value of the velocity of the particle at the position $x$ is
\begin{equation}\label{vharm}
v=\sqrt{k\left(x_0^2-x^2\right)}
\end{equation}
and the rescaled PDF of the return phase is
\begin{equation}  \label{I2k}
\rho_2(x)=\int_{|x|}^\infty dx_0 \frac{1}{\sqrt{k\left(x_0^2-x^2\right)}}\int_0^\infty dt\; p(x_0|t) \psi(t)\,.
\end{equation}
The return time attains the value
\begin{equation}\label{tretspring}
\left\langle t_{\rm ret}\right\rangle=\frac{\pi }{2\sqrt{k}}
\end{equation} 
which does not depend on the position $x_0$ in the beginning of the return motion in contrast to the case of return with constant speed (Eq.~\ref{ttretv}) and constant acceleration (Eq.~\ref{ttreta}). The corresponding trajectories are shown at Fig.~\ref{Gxyza} with a blue line.

\begin{figure*}[ht]
\centerline{\includegraphics[width=0.98\columnwidth]{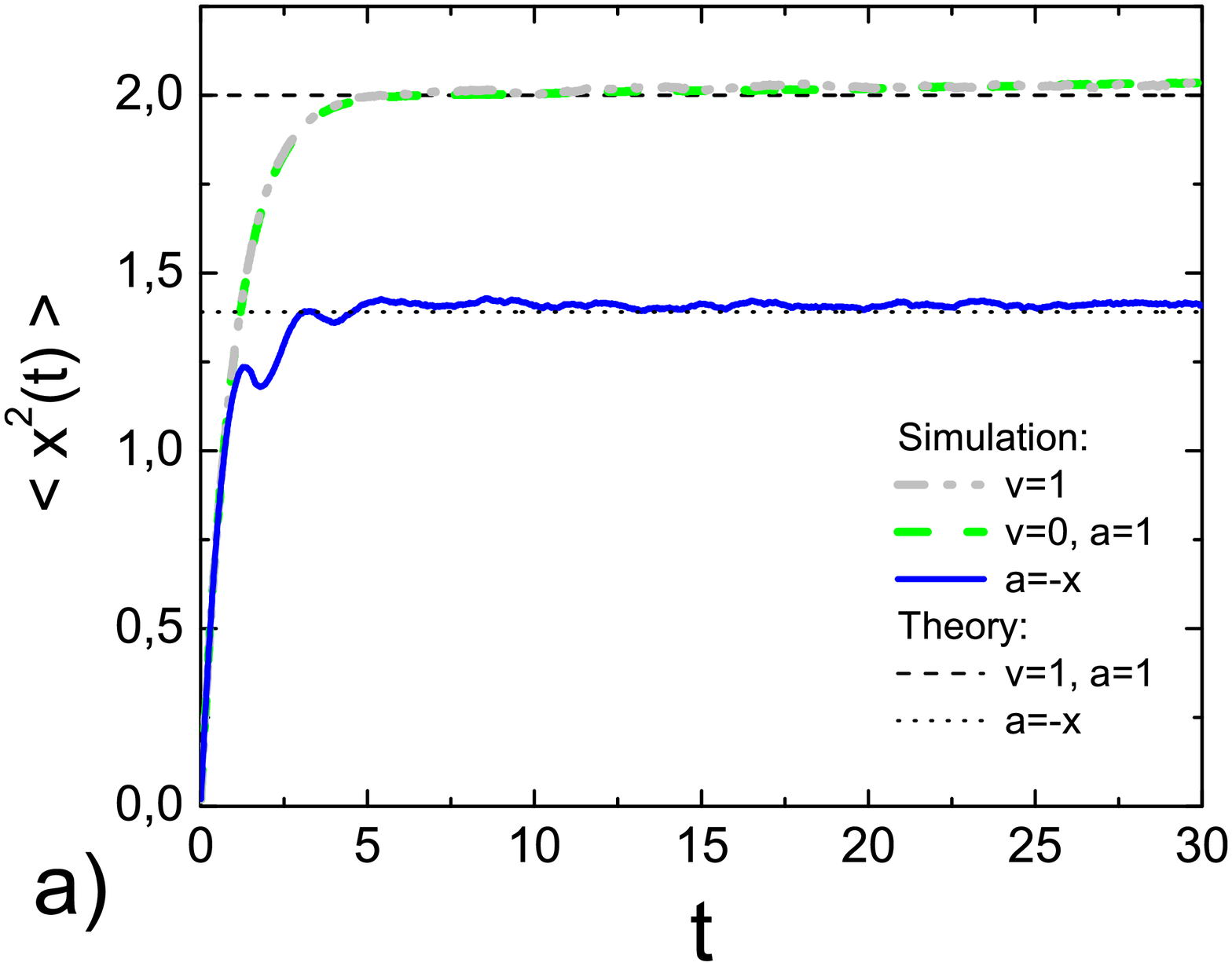}\includegraphics[width=1.06\columnwidth]{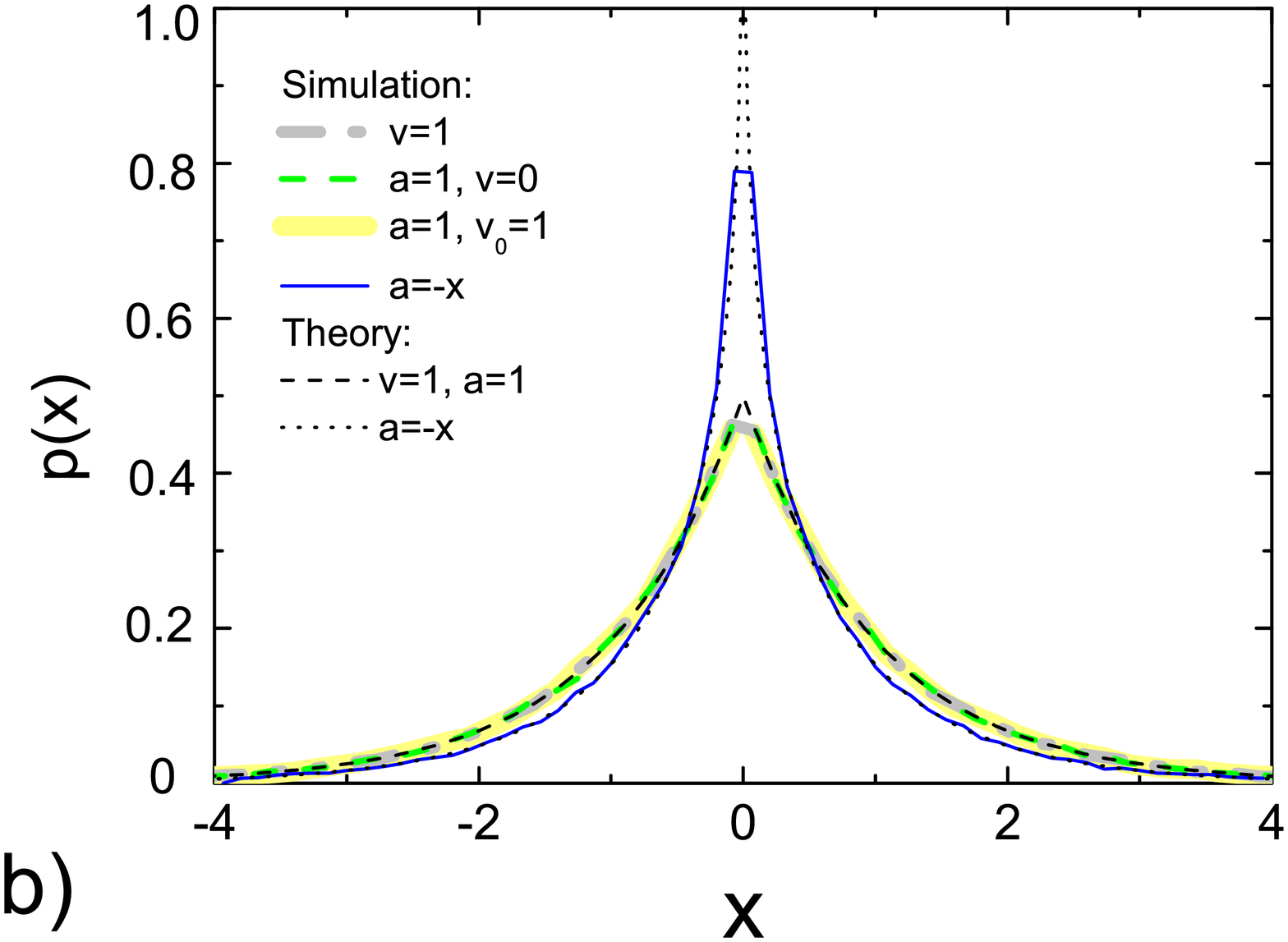}}
\caption{Mean-squared displacement (a) and probability density function (b) for the exponential resetting
with $D=1$, $r=1$. The dashed black line in panel (a) shows the analytical asymptotic result for the MSD, Eq.~(\ref{x2exp}), the dashed black line
in panel (b) shows the result for the PDF from  Eq.~(\ref{pdfexp}). }
\label{Gexp}
\end{figure*} 

\subsection{Computer simulations} 

The analytical predictions for MSD and PDF, as well as for the mean hitting time, discussed later, are compared with 
numerical simulations directly following from the discretization of Langevin equations in the motion phase and of the equation 
of motion for the return at a constant speed. The time axis is discretized with the step $dt=t_{i+1}-t_i$, and the time of 
the first resetting event is generated according to its probability density $\psi(t)$. For the deterministic resetting this 
resetting time is fixed. During the displacement phase the particle performs stochastic motion according to a finite-difference 
analogue of the Langevin equation
\begin{equation}\label{lattice}
x_{i+1}=x_i+\xi_i\sqrt{2Ddt}.
\end{equation}
Here $x_i=x(t_i)$ is the coordinate of the particle at time $t_i$, and
$\xi_i$ is the random number distributed according to a standard normal distribution generated using the Box-Muller transform. 
When the resetting event occurs, the particle starts moving to the origin according to either Eq.~(\ref{xvret}) or Eq.~(\ref{xxxret}) or Eq.~(\ref{xharm}) depending on the type of the return motion. When the particle crosses the origin, the time of the next resetting event is generated, and the particle starts performing stochastic motion until this resetting event. 
All simulations are performed with $N=10^5$ particles. The simulation of the trajectories of the particles is shown in Fig.~\ref{fig:trajectories} for the return at constant speed and in Fig.~\ref{Gxyza} for the return at constant acceleration and under the action of the harmonic force.

\begin{figure*}[htbp]
\centerline{\includegraphics[width=0.98\columnwidth]{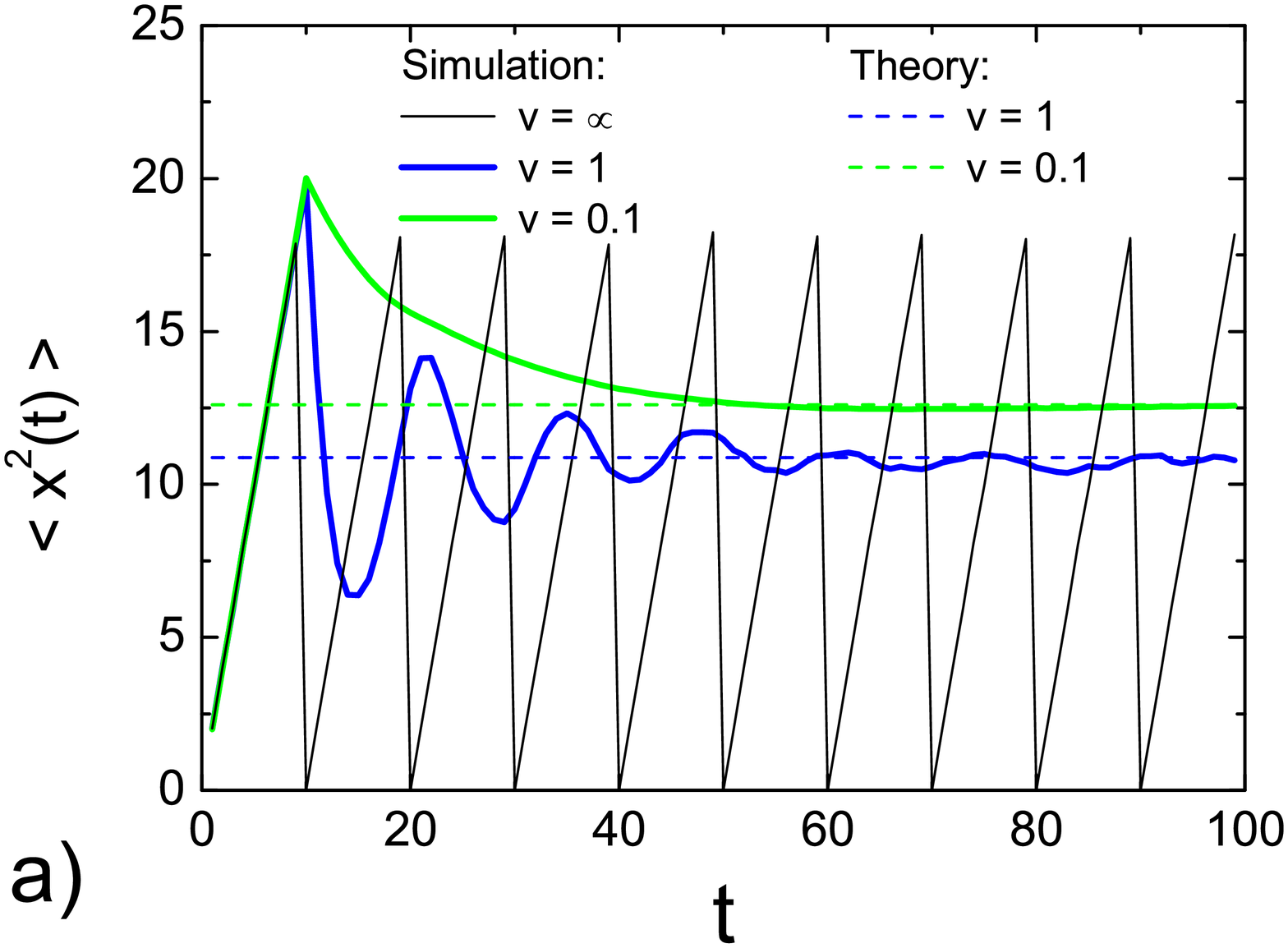}\includegraphics[width=0.98\columnwidth]{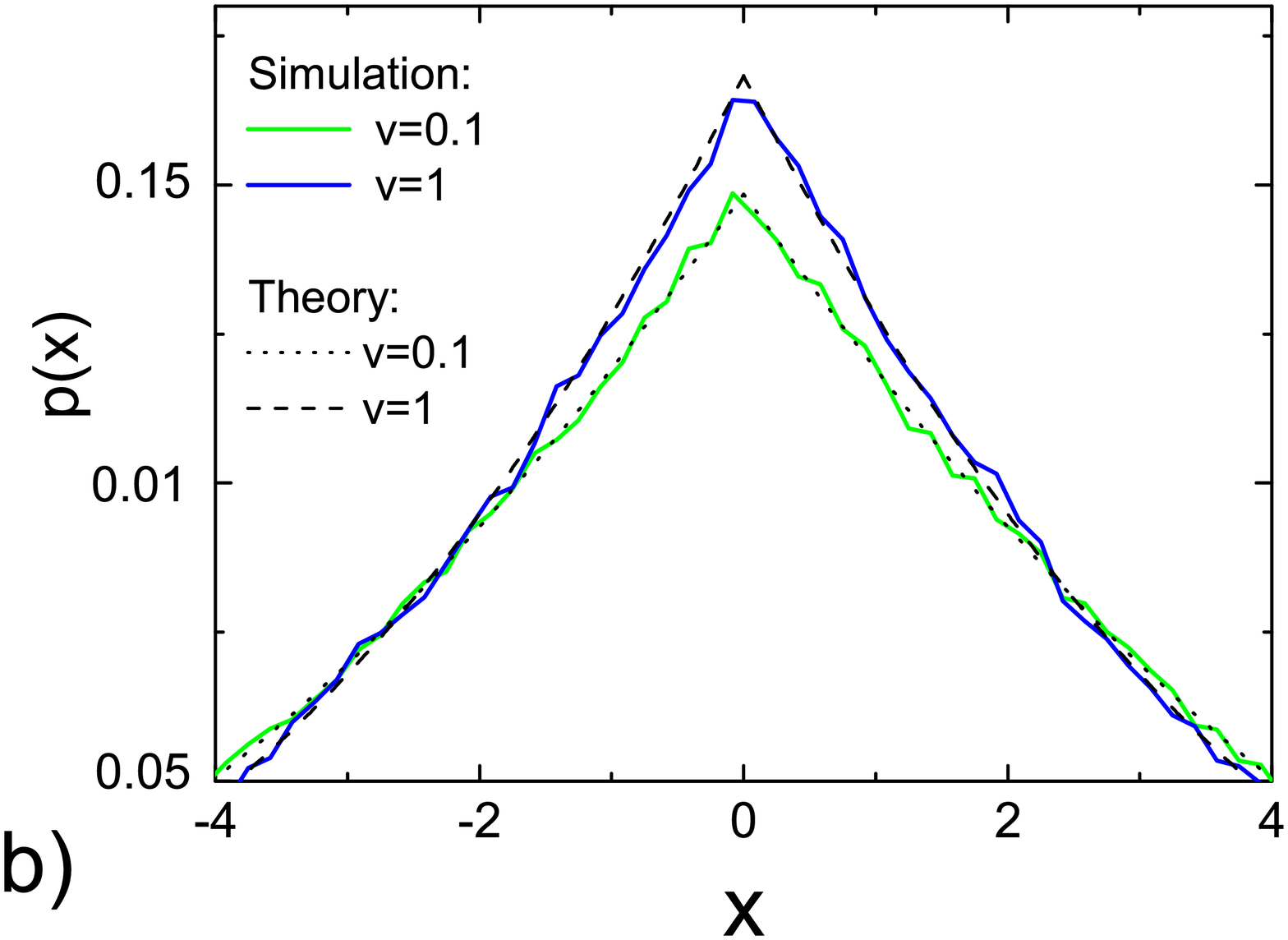}}
\caption{ (a) Mean-squared displacement as a function of time and (b) stationary probability density function at $t=100$ for the case of fixed-time resetting and the return at constant speed. The dashed lines in panel (a) show the analytical asymptotic results for the MSD, Eq.~(\ref{x2det}), the dotted and dashed black lines 
in panel (b) are the results for the PDF from  Eq.~(\ref{pdfdelta}). The resetting time is $t_r=10$, the diffusion coefficient $D=1$.}
\label{Gdet}
\end{figure*}

\section{Diffusion with exponential resetting.}

Let us consider standard Brownian motion with PDF given by Eq.~(\ref{pgau}) an exponential waiting time distribution of the resetting events:
\begin{equation}\label{pexp}
\psi(t)=r\exp(-rt) .
\end{equation}
The survival probability is $\Psi(t)=e^{- rt}$, the average location at the resetting event is equal to $\langle |x_0| \rangle=\sqrt{D /r}$ and the mean resetting time is
\begin{equation}\label{tresexp}
\langle t_{\mathrm{res}} \rangle = \frac{1}{r}.
\end{equation}
The rescaled PDF of the displacement phase $\rho_1(x)$ is given by the Laplace distribution:
\begin{equation}\label{III1}
\rho_1(x)=\frac{1}{2\sqrt{Dr}}\exp\left(-\frac{ |x|}{\sqrt{D/r}}\right) .
\end{equation}

\subsection{Return at a constant speed}
If the return velocity does not change during the return motion or depends only coordinate $x$ and does not depend on $x_0$, $\rho_2$ is also given by the Laplace distribution
\begin{equation}\label{III2}
\rho_2(x)=\frac{1}{2v}\exp\left(-\frac{ |x| }{\sqrt{D/r}}\right)\,,\end{equation}
but with a different prefactor compared to the displacement part of the PDF (Eq.~\ref{III1}).
The condition of the stationarity of the ratio of two integrals given by Eq.~(\ref{I1I2C}) is fulfilled with $C=\sqrt{Dr}/v$. The PDF becomes completely invariant with respect to the return process and is the same as PDF of normal diffusion with instantaneous exponential resetting \cite{EvansMajumdar}:
\begin{equation}
 P(x)=  \frac{1}{2 \sqrt{D /r}} \exp\left(-\sqrt{\frac{r}{D}}|x|\right)\,. \label{pdfexp}
 \end{equation}
The same effect has been obtained independently in \cite{shlomi1,shlomi2}, using a different approach. The MSD can be derived according to Eq.~(\ref{msdpdf})
\begin{equation}\label{x2exp}
\left\langle x^2\right\rangle = \frac{2D}{r}\,
\end{equation}
also independently of the velocity of resetting. The analytical and numerical results for the MSD and the PDF for Brownian motion with exponential resetting are depicted and compared at Fig.~\ref{Gexp}. Fig.~\ref{Gexp}b confirms the result that for the case of the Brownian motion with exponential resetting, the models with instantaneous return and with the return at a constant speed lead to the same stationary probability densities of the coordinate. The MSD rapidly tends to a steady state value given by Eq.~(\ref{x2exp}) (Fig.~\ref{Gexp}a).

\subsection{Return at $v=v(x_0-x)$}
Let us assume that the velocity of particle depends only on the displacement from the initial position $x_0$: $v=v\left(x_0-x\right)$. In this case 
\begin{equation}
\rho_2(x)=\frac{\hat{C}}{2}e^{-\sqrt{\frac{r}{D}}|x|}\,,
\end{equation}
where
\begin{equation}
\hat{C}=\int_0^{\infty}\frac{dz}{2v(z)}\sqrt{\frac{r}{D}}e^{-z\sqrt{\frac{r}{D}}}\,.
\end{equation}
Therefore, the fraction $\rho_2(x)/\rho_1(x)$ does not depend on the location $x$, and the PDF has the same form as for the instantaneous resetting (Eq.~\ref{pdfexp}). 

\subsection{Return at a constant acceleration}

For the return with a constant acceleration $a$ with zero initial velocity $v_0=0$ we obtain
\begin{equation}\label{Xi2}
\rho_2(x)=\sqrt{\frac{\pi}{8a}}\left(\frac{r}{D}\right)^{\frac{1}{4}}\exp\left(-\frac{|x|}{\sqrt{D/r}}\right)
\end{equation}
The condition, Eq.~(\ref{I1I2C}) is fulfilled again with 
\begin{equation}\label{C1a}
C=\sqrt{\frac{\pi}{2a}}D^{\frac{1}{4}}r^{\frac{3}{4}}\,.
\end{equation}
The average return time can be obtained according to Eq.~(\ref{tretI}): $\langle t_{\rm ret}\rangle=\sqrt{\frac{\pi}{2|a|}}\left(\frac{D}{r}\right)^{\frac{1}{4}}$.
The PDF
is again independent on the particular value of the acceleration of the particles and has the same form of the Laplace distribution as PDF for instantaneous resetting (Eq.~\ref{pdfexp}). The MSD is the same as well, as shown in Fig.~\ref{Gexp}a.  If the return motion occurs with constant acceleration $a$, but the initial return velocity $v_0$ is non-zero, the return velocity $v$ is also the function only of the difference $x_0-x$: $v=2v_0-\sqrt{v_0^2+2a(x_0-x)}$.
Therefore, the final form of the PDF again remains the same as in the case of the instantaneous resetting (See Fig.~\ref{Gexp}b).

\begin{figure*}[htbp]
\centerline{\includegraphics[width=0.98\columnwidth]{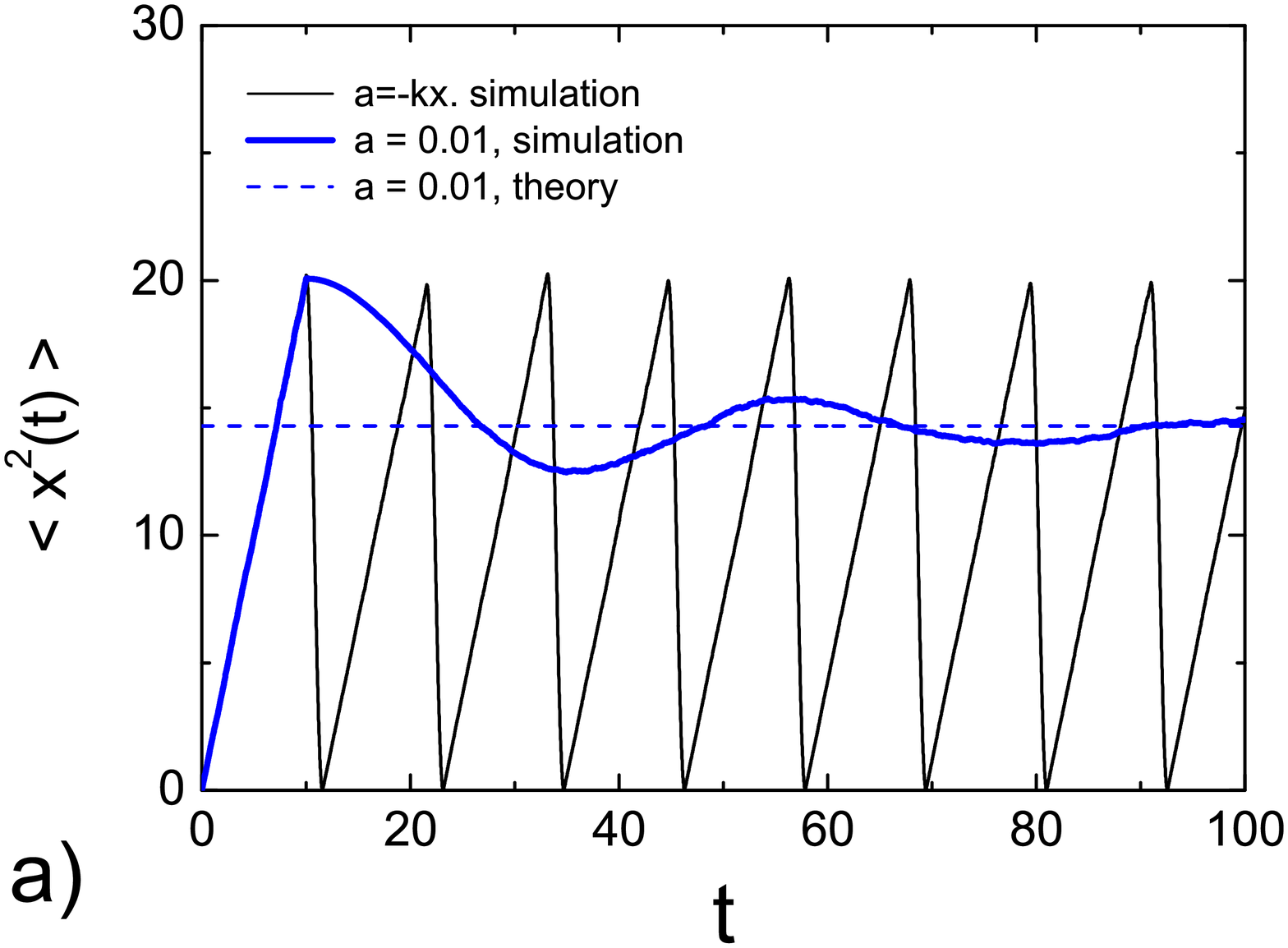}\includegraphics[width=0.98\columnwidth]{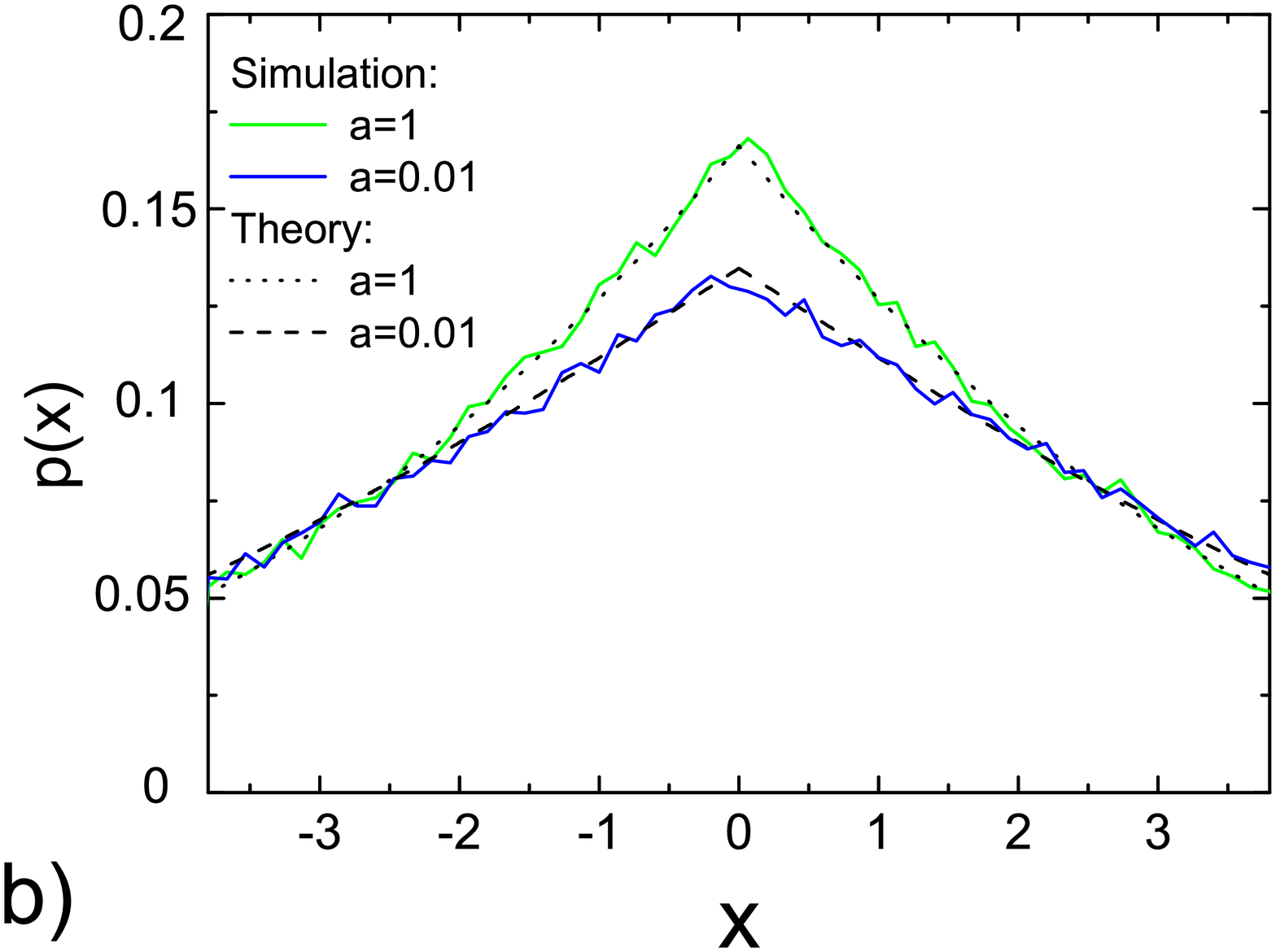}}
\caption{ (a) Mean-squared displacement as a function of time (for the return at constant acceleration and with harmonic motion) and (b) stationary probability density function  at $t=100$ for the case of fixed-time resetting and return with constant acceleration. The dashed lines in panel (a) show the analytical asymptotic results for the MSD, Eq.~(\ref{x2det}), the dotted and dashed black lines 
in panel (b) are the analytical results for the PDF obtained by incorporating Eqs.~(\ref{Ii1}) and~(\ref{J2a}) into Eq.~(\ref{pdfdeta}). The resetting time is $t_r=10$, the diffusion coefficient $D=1$.}
\label{Gdeta}
\end{figure*}

\subsection{Return under the action of a harmonic force}
If the velocity explicitly depends on $x$ and $x_0$, then such an invariance does not hold. Let us consider the return according to the harmonic motion, Eq.~(\ref{xharm}). The rescaled displacement PDF is in this case equal to
\begin{equation}\rho_2(x)=\frac{1}{2}\sqrt{\frac{r}{kD}}K_0\left(\frac{\left|x\right|}{\sqrt{D/r}}\right)\end{equation}
and the PDF is not the same as for the instantaneous resetting (blue line at Fig.~\ref{Gexp}b):
\begin{equation}\label{pspring}
P(x)=\frac{\frac{1}{\sqrt{Dr}}\exp\left(-\frac{ |x|}{\sqrt{D/r}}\right)+\sqrt{\frac{r}{kD}}K_0\left(\frac{\left|x\right|}{\sqrt{D/r}}\right)}{\frac{2}{r}+\frac{\pi}{\sqrt{k}}}\,.
\end{equation}
The MSD can be obtained according to Eq.~(\ref{msdpdf}):
\begin{equation}\label{x2spring}
\langle x^2\rangle=\frac{4D\sqrt{k}+\pi Dr}{2r\sqrt{k}+\pi r^2}\,.
\end{equation}
The simulation data for MSD are shown at Fig.~\ref{Gexp}a as a solid blue line. The MSD rapidly tends to the steady state given by Eq.~(\ref{x2spring}).

\section{Diffusion with deterministic resetting.}

Let us now consider normal diffusion with PDF given by Eq.~(\ref{pgau}) under resetting at a constant pace:
\begin{equation}\label{pdelta}
\psi(t)=\delta(t-t_r).
\end{equation}
The survival probability for resetting events 
is now $\Psi(t)=\Theta(t_r-t)$. The mean duration of the displacement phase is $\langle t_{\mathrm{res}} \rangle =t_r$ and the average location at the resetting events $\langle |x_0 | \rangle=\sqrt{4Dt_r/\pi}$. The first integral is
\begin{equation}\label{Ii1}
\rho_1=\sqrt{\frac{t_r}{\pi
D}}\exp\left(-\frac{x^2}{4Dt_r}\right)-\frac{\left|x\right|}{2D}
\mbox{erfc}\left(\frac{\left|x\right|}{\sqrt{4Dt_r}}\right)
\end{equation}

\subsection{Return at a constant speed}
For the return at constant speed the rescaled return PDF takes the form
\begin{equation}
\rho_2=\frac{1}{2v}\mbox{erfc}\left(\frac{\left|x\right|}{\sqrt{4Dt_r}}\right), \label{Jj2}
\end{equation}
where $\rm erfc(x)$ is the complementary error function. In such a way the PDF attains the form 
\begin{equation}\label{pdfdelta}
  P(x)=\frac{\rho_1(x)+\rho_2(x)}{t_r+\frac{1}{v}\sqrt{\frac{4Dt_r}{\pi}}} .
\end{equation}
with $\rho_1(x)$ given by Eq.~(\ref{Ii1}) and $\rho_2(x)$ given by Eq.~(\ref{Jj2}).
The MSD, given by Eq.~(\ref{msdpdf}), reads
\begin{equation}\label{x2det}
\left\langle x^2\right\rangle = Dt_r\left(1+\frac{\frac{2}{3} \sqrt{Dt_r}}{t_r v\sqrt{\pi}+ 2\sqrt{Dt_r}}\right)    \,.
\end{equation}
The quotient of $\rho_2$ given by Eq.~(\ref{Jj2}) and $\rho_1$ given by Eq.~(\ref{Ii1}) is not constant any more and now explicitly depends on $x$. Consequently, the PDF and the MSD are not invariant with respect to the return velocity, as found for the case of exponential resetting. The results for the PDF and the following MSD are represented in Fig.~\ref{Gdet} by dashed lines.  The oscillations of the MSD are observed. For instantaneous return these oscillations persist indefinitely. 
Note that our Eq.~(\ref{x2det}), by construction, only predicts the time-averaged MSD (and therefore gives a constant value $\left\langle x^2\right\rangle = Dt_r$) but not its oscillations
at the pace dictated by the resetting process.
For the return at finite velocity the oscillations are damped: if the velocity is high, the MSD performs several perceptible oscillations when approaching a constant value given by Eq.~(\ref{x2det}). 
At small velocities the MSD approaches this value monotonically after the initial overshooting. The final value of the MSD at long times is correctly reproduced by Eq.~(\ref{x2det}). 

\subsection{Return at a constant acceleration}
In the case of the return with constant acceleration
\begin{equation}\label{J2a}
\rho_2(x)=\sqrt{\frac{|x|}{16\pi a Dt_r}}e^{-\frac{x^2}{8Dt_r}}K_{1/4}\left(\frac{x^2}{32Dt_r}\right)\,,
\end{equation}
where $K_{1/4}$ is the modified Bessel function of second kind.
The average return time is
\begin{equation}\label{tretdeta}
\langle t_{\rm ret}\rangle=\sqrt{\frac{2}{\pi a}}\left(4Dt_r\right)^{\frac14}\Gamma\left(\frac34\right)\,.
\end{equation}
The PDF attains the form
\begin{equation}\label{pdfdeta}
  P(x)=\frac{\rho_1(x)+\rho_2(x)}{t_r+\sqrt{\frac{2}{\pi a}}\left(4Dt_r\right)^{\frac14}\Gamma\left(\frac34\right)} ,
\end{equation}
with $\rho_1(x)$ given by Eq.~(\ref{Ii1}) and $\rho_2(x)$ given by Eq.~(\ref{J2a}). The PDF is depicted at Fig.~\ref{Gdeta}b. It is again not invariant with respect to the acceleration, analogously to the case of the return with constant velocity. The MSD can be obtained by numerical integraton of the PDF given by Eq.~(\ref{pdfdeta}) according to Eq.~(\ref{msdpdf}).

\subsection{Return under the action of a harmonic force}
For the simple harmonic return motion we obtain
\begin{equation}\rho_2(x)=\frac{1}{4\sqrt{\pi kDt_r}}K_0\left(\frac{x^2}{8Dt_r}\right)\exp{\left(-\frac{x^2}{8Dt_r}\right)}\,.
\end{equation}
The return time is given in terms of Eq.~(\ref{tretspring}). 
Interestingly, in contrast to the return with constant speed and acceleration, the average return time does not depend on the fixed resetting time $t_r$ and depends only on strength of the spring $k$. Consequently, the duration of the run remains always the same. The MSD oscillates as in the case of instantaneous deterministic resetting (see Fig.~\ref{Gdeta}b). The oscillations of MSD are not damped due to the fixed duration of the run. Both MSD and PDF never attain a stationary value.

\section{The invariance of PDF with respect to the return velocity}
\label{sec:Invariance}

Let us summarize the findings of the previous section. We have observed that the the PDF under exponential resetting is invariant with respect to return speed and acceleration for the case of normal diffusion \cite{shlomi1,shlomi2}. On the contrary, for resetting at a constant pace such invariance does not hold. Therefore the interesting question arise: what kind of conditions are necessary for such an invariance to take place. Below we consider only the return at constant speed. The case of the return at a constant acceleration can be treated analogously.

For the return at a constant speed $|v(x;x_0)| = v$ we immediately get from Eq.~(\ref{I2vv})
\begin{eqnarray}  \label{I2v}
&&\rho_2(x)=  \frac{\hat{\rho}_2(x)}{v}\,, \\
&&\hat{\rho}_2(x)=\int_{|x|}^\infty dx_0 \int_0^\infty dt\; p(x_0|t) \psi(t)\,. \label{J2}
\end{eqnarray}
The PDF given by Eq.~(\ref{mixture}) takes the form
\begin{equation}\label{pgeneral}
 P(x) = \frac{\rho_1(x) + v^{-1}\hat{\rho}_2(x) }{\langle t_{\rm res} \rangle + v^{-1} \langle |x_0 | \rangle} .
\end{equation}
We note that 
\begin{equation}
 \int_0^\infty \hat{\rho}_2(x) dx = \langle |x_0 | \rangle.
 \label{eq:IntJ2}
\end{equation}

Let us start from Eq.~(\ref{pgeneral}) and require $P(x)$ on the l.h.s. to be independent on $v$.
Taking the derivative of the r.h.s. with respect to $v^{-1}$ and setting this derivative equal to zero we get
\begin{equation}
 \frac{\hat{\rho}_2(x)}{\rho_1(x)} = \frac{\langle |x_0 | \rangle}{\langle t_{\rm res} \rangle}.
 \label{eq:proportion}
\end{equation}
The right hand side of this equation does not depend on $x$ and depends only on the properties of the transport phase, but not on the return speed. 

By virtue of Eqs.~(\ref{eq:IntI1}) and (\ref{eq:IntJ2}) we can see that if the quotient $\hat{\rho}_2(x)/\rho_1(x)$ is independent on $x$, it is equal to the right hand side of the Eq.~(\ref{eq:proportion}) automatically. Therefore the only condition for the velocity independence is
\begin{equation}\label{eq:rel2int}
\frac{\hat{\rho}_2(x)}{\rho_1(x)} = C.
\end{equation}
where $C=\rm const$. Hence, we show that the condition given by Eq.~(\ref{I1I2C}) is not only sufficient, but also the necessary condition for the invariance of the PDF with respect to the return velocity.

Eq.~(\ref{eq:rel2int}), together with Eqs.~(\ref{I1}) and (\ref{J2}), defines an awkward equation involving different integrals of $\psi(t)$ and 
$p(x|t)$. This relation can be reduced to a functional relation between only two functions. Let us fix some $x >0$ (the case $x<0$ follows by symmetry).
Interchanging the order of integration in $t$ and $t_{\mathrm{res}}$ in Eq.~(\ref{I1}) we can express the first integral via the survival probability
in a displacement phase $\Psi(t) = \int_t^\infty \psi(t') dt'$ and the function $F(x |t) = \int_x^\infty p(x'|t) dx'$:
\begin{eqnarray}
\rho_1(x) &=&  \int_0^\infty dt p(x|t) \int_t^\infty \psi(t_{\mathrm{res}}) d t_{\mathrm{res}} \nonumber \\
&=& -\int_0^{\infty} \left[\frac{\partial }{\partial x} F(x|t) \right] \Psi(t) dt . \label{I11}
\end{eqnarray}
In Eq.~(\ref{J2}) we first integrate over $x_0$ and then perform partial integration in time:
\begin{equation}\label{J2new}
\hat{\rho}_2(x) =  \int_0^\infty \left[ \frac{\partial}{\partial t} F(x|t) \right]\Psi(t)dt.
\end{equation}
Substituting Eq.~(\ref{I11}) and Eq.~(\ref{J2new}) into Eq.~(\ref{eq:rel2int}), one can formulate the condition of the invariance of the PDF with respect to the return velocity as an orthogonality relation
\begin{equation}
 \int_0^\infty \left[C\frac{\partial }{\partial x} F(x|t) +  \frac{\partial}{\partial t} F(x|t)\right] \Psi(t) dt = 0.
 \label{eq:Ort}
\end{equation}
This means that for given $F(x | t)$ the function $\Psi(t)$ must be orthogonal to the whole family of functions of time $f_x(t) = C \frac{\partial }{\partial x} F(x|t) +  \frac{\partial }{\partial t} F(x|t)$ defined on $[0,\infty)$ and parametrized by $x$ for some particular value of $C$. The fact that $\Psi(t)$ is a survival probability poses considerable restrictions on the possible solution: 
it has to be a  non-negative, monotonically decaying function of $t$ with $\Psi(0)=1$. When any of these restrictions is violated, no physical solution $\Psi(t)$ exists. 

Eq.~(\ref{eq:Ort}) is quite tricky. Thus, for a Gaussian PDF $p(x|t)$ it is easy to show that it is solved by $\Psi(t) = \exp(-r t)$ and find $C = \sqrt{Dr}$.
However, we were not able to show that this exponential solution is unique. 
In the present work we do not concentrate on general pathway to solution of Eq.~(\ref{eq:Ort}), but, using the notation adopted in Eqs.~(\ref{I11}) and (\ref{J2new}),
show two simple (somewhat degenerate) examples corresponding to ballistic motion during the displacement phase. These examples 
demonstrate that there exist situations, in which the PDF $P(x)$ is invariant under return speed for any resetting waiting time PDF, and also situations, in which 
no waiting time PDF can be found, under which such an invariance takes place. 

\subsection{Ballistic motion during the displacement phase with constant velocity}

Let us consider the motion taking place at a constant velocity with the probability density function
\begin{equation}
 p(x|t) = \delta(x-v_0 t) .
\end{equation}
The integrals are equal to
\begin{eqnarray}
 \rho_1(x) =  \frac{1}{v_0}\Psi\left(\frac{x}{v_0}\right),\\
 \hat{\rho}_2(x) =  \Psi\left(\frac{x}{v_0}\right) .
\end{eqnarray}
 Eq.(\ref{eq:rel2int}) is fulfilled for $C = v_0$ for \textit{any} $\Psi(t)$: this process is invariant with respect to the return velocity for any distribution of the resetting times. 
 The same consideration is valid for the symmetric distribution 
 \begin{equation}
 p(x|t) = \frac12\delta(x-v_0 t)+\frac12\delta(x+v_0 t).
\end{equation}

\subsection{Ballistic motion during the displacement phase with random velocity}

Now we demonstrate an example of $p(x|t)$ for which no matching $\Psi(t)$ can be found. We namely consider a variant of the previous model, but now with the transport velocity 
which is randomly chosen from the interval $[0,V]$ at each run. The probability density in this case attains the form
\begin{equation}
 p(x|t) = (V t)^{-1} \Theta(V t - x).
\end{equation}
Our integrals are 
\begin{eqnarray}
 \rho_1(x) =  \frac{1}{V}\int_{x/V}^\infty \frac{\Psi(t)}{t} dt , \\
 \hat{\rho}_2(x) =  \frac{1}{V}\int_{x/V}^\infty \frac{x}{t} \frac{\Psi(t)}{t} dt .
\end{eqnarray}
The proportionality of these two integrals means that 
\begin{equation}
\frac{C}{V} \int_{x/V}^\infty \frac{\Psi(t)}{t} dt =  \int_{x/V}^\infty \frac{x}{V t} \frac{\Psi(t)}{t} dt .
\end{equation}
 Introducing a new constant $c= V/C$, and new independent and dependent variables $y = x/V$ and $\eta(t) = \Psi(t)/t$  we get
\begin{equation}
 \int_y^\infty \eta(t) dt = c y \int_y^\infty \frac{\eta(t)}{t}  dt. \end{equation}
Differentiating both parts with respect to $y$ twice we get
\begin{equation}
 \frac{c-1}{c} \eta^{\prime}(y) = -\frac{\eta(y)}{y}.
\end{equation}
The solution of this equation is $\eta(y) = A y^{- \frac{c}{c-1}}$
so that
\begin{equation}
 \Psi(t) = B t^{-\frac{1}{c-1}}.
\end{equation}
This function however cannot be a survival probability because if it decays at infinity (which implies $c>1$)
it diverges at zero and therefore $\Psi(0) = 1$ cannot be fulfilled under any circumstances.

\begin{figure}[ht] 
\includegraphics[width=0.85\columnwidth]{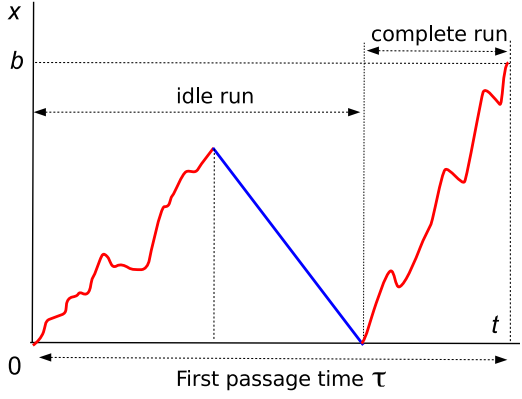}
 \caption{The target is located at $x=b$. The particle starts its motion at the origin $x=0$. If a target is found during the run, the run is complete. Otherwise, the run is termed idle. \label{Gtau}} 
\end{figure}

\section{Mean first passage time \label{sec:MFPT}} 
When turning to mean first hitting (or first passage) properties of the process, we will consider the situation when 
the hitting only can take place during the displacement phase. In one dimension this is trivially the case, since 
the interval of the $x$-axis from the origin to $x_0$ was already covered by the trajectory of the displacement process. 
Therefore, if there were any targets on this interval, they should already be found during the displacement phase. The situation in higher dimensions is richer. 
Depending on the properties of the searcher, one can assume that, parallel 
to the 1d situation, the target can only be found during the displacement phase, or during both, displacement and return,
phases. We do not consider these high-dimensional situations in the present work. Our approach to the problem will be 
similar to the one used in \cite{SokChech}. 

Let us denote the run as complete, if the target is found and idle, if the target is not found (Fig.~\ref{Gtau}). 
The hitting of the target may occur during the first run, during the second run (the first one is idle and the second is complete as shown at Fig.~\ref{Gtau}) or the third run (two first runs are idle, i.e. finished), etc., and the probability $h(t)$ to reach a target at time $t$ attains the following form:
\begin{eqnarray}
 &&  h(t) = \omega(t) + \int_0^t \phi(t') \omega(t-t') dt' \\
 && \qquad +  \int_0^t  \int_0^{t'} \phi(t'') \phi(t') \omega(t-t'-t'') dt'dt'' + ... \;.\nonumber
\end{eqnarray}

\begin{figure*}[ht]
\centerline{\includegraphics[width=\columnwidth]{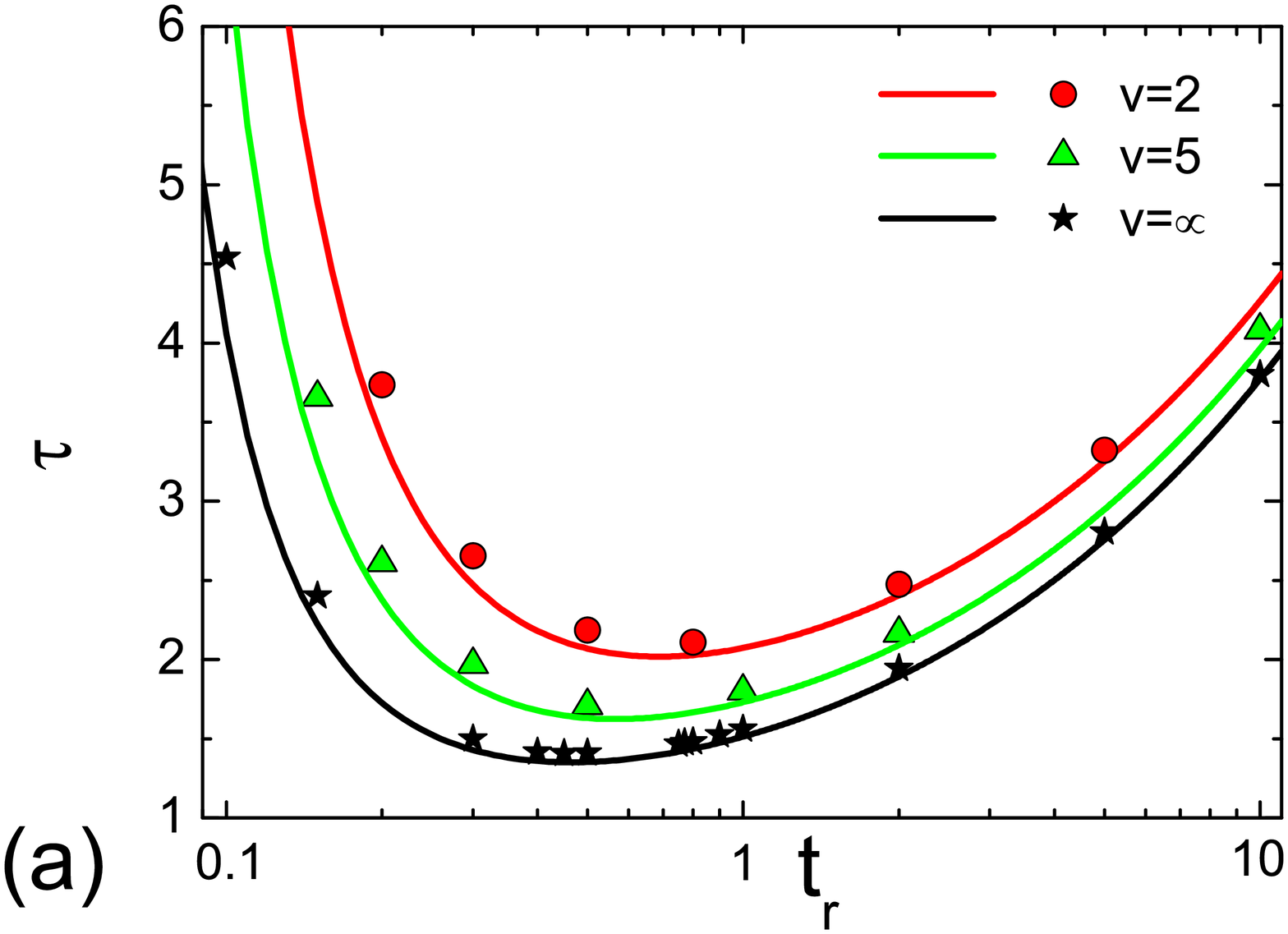}\includegraphics[width=\columnwidth]{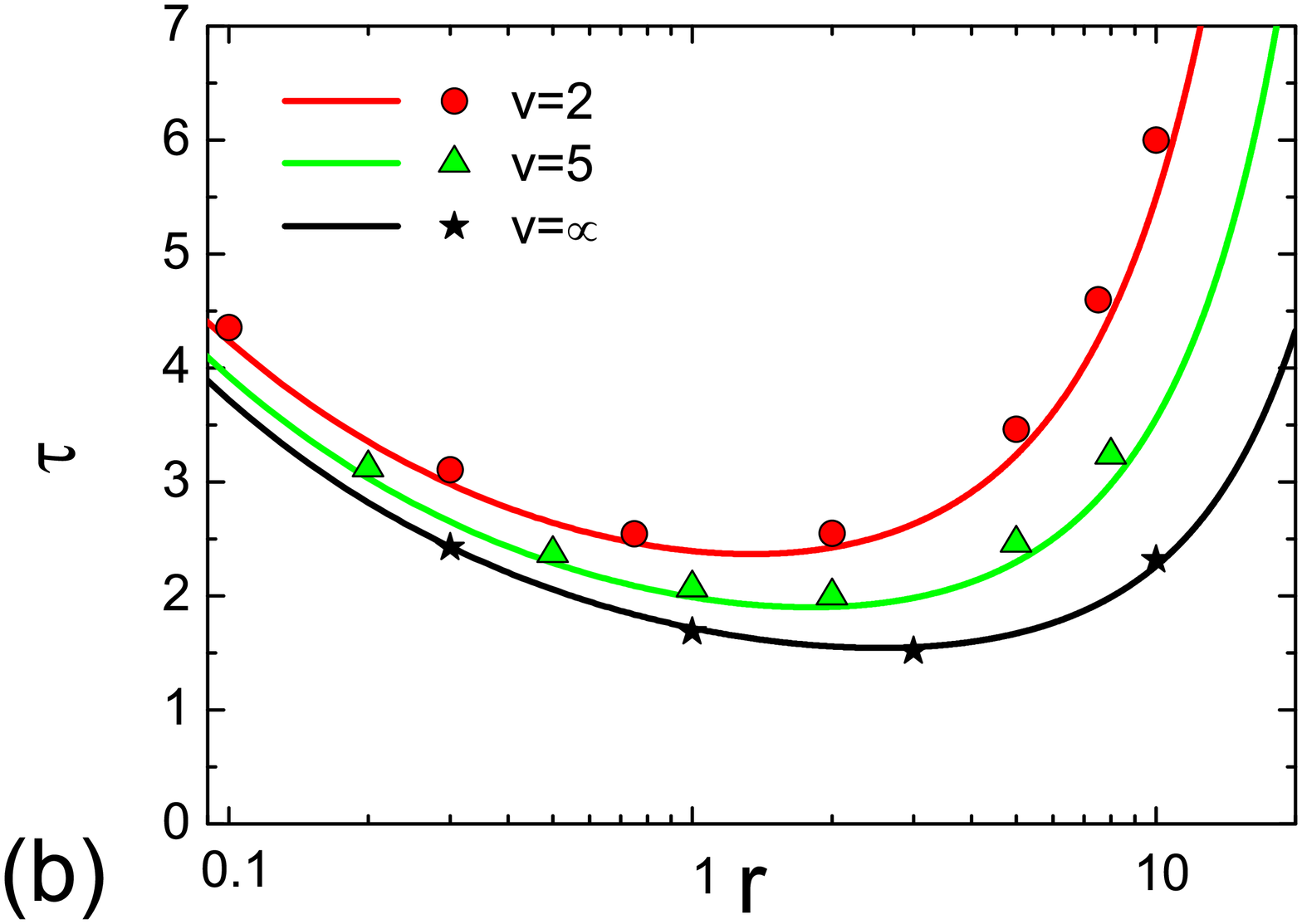}}
\caption{Mean first passage times for the return at constant speed for (a)
$\psi(t)=\delta(t-t_r)$, and (b) $\psi(t)=re^{-rt}$. Symbols correspond to the simulation data, solid lines -- to
the analytical result, Eq.~(\ref{tauall1}) with $\tau_0$ and $l_v$ given by Eqs.~(\ref{tau0det}) and (\ref{l0det}) for the resetting at constant pace and Eqs.~(\ref{tau0exp}) and (\ref{l0exp}) for the exponential resetting. The target is located at $b=1$. $v=\infty$ corresponds to the case of instantaneous return.}
\label{Gfpt}
\end{figure*}

Here $\omega(t)$ is the probability density of hitting a target at time $t$ after starting a (complete) run, and
$\phi(t)$ is a probability density of finishing an idle run at time $t$ after its beginning. Since the series has a 
form of a sum of multiple convolutions of $\omega(t)$ and $\phi(t)$, this can be easily evaluated in the Laplace domain:
\begin{equation}
\tilde{h}(s) = \tilde{\omega}(s) + \tilde{\phi}(s) \tilde{\omega}(s) + \tilde{\phi}^2(s) \tilde{\omega}(s) + ... =
\frac{\tilde{\omega}(s)}{1-\tilde{\phi}(s)}.
\end{equation}
Here $\tilde{\omega}(s)=\int_0^{\infty}dt e^{-ts}\omega(t)$ is the Laplace transform of $\omega(t)$.  The MFPT $\tau$ can be obtained in the following way:
\begin{equation}
\tau = - \left. \frac{d}{ds} \tilde{h}(s) \right|_{s=0} = - \frac{\tilde{\omega}'(0)}{1- \tilde{\phi}(0)} - \frac{\tilde{\omega}(0)
\tilde{\phi}'(0)}{(1 - \tilde{\phi}(0))^2}\,. \label{taudef}
\end{equation}
Let us introduce the probability $P_c=\tilde{\omega}(0)$ that the run is complete or that the
target has been found during the run. $P_i=\tilde{\phi}(0)$ is the probability that a
run is finished (idle) and the target has not been found during this run. 
Each run is either complete or idle: $P_c + P_i = 1$, therefore we get from Eq.~(\ref{taudef}):
\begin{equation}
\tau = - \left(\tilde{\omega}'(0)+\tilde{\phi}'(0)\right)/P_c.
\label{eq:MFPT}
\end{equation}

Taking the derivative of the Laplace transform of $\omega$ one can get $\tilde{\omega}'(0) = - \int_0^\infty t \omega(t) dt$. 
Dividing this by $P_c$ we obtain the mean
duration of a complete run: $\langle t_{\mathrm{hit}} \rangle=-\tilde{\omega}'(0)/ P_c $. Analogously, for the idle run $\langle t_{\mathrm{run,b}} \rangle=-\tilde{\phi}'(0)/ P_i$. Taking into account that $\langle t_{\mathrm{run,b}}\rangle = \langle t_{\mathrm{res}}\rangle + \langle t_{\mathrm{ret,b}}\rangle$, 
we get 
\begin{equation}\label{taunorm}
\tau = \frac{\langle t_{\mathrm{hit}}\rangle P_c+ \langle t_{\mathrm{res}} \rangle P_i + \langle t_{\mathrm{ret,b}} \rangle P_i}{P_c}. 
\end{equation} 
Note that the average run and return times, $\langle t_{\mathrm{run,b}}\rangle$ and $\langle t_{\mathrm{ret,b}}\rangle$, differ from the quantities given by Eq.~(\ref{trun}) and Eq.~(\ref{tret}), correspondingly, because now the motion of the particle is restricted by the adsorbing target, located at $x=b$.

Let us take into account that $P_i$ is the probability that $t_{\mathrm{res}}<t_{\mathrm{hit}}$ and $P_c$ is the probability that the time $t_{\mathrm{hit}}<t_{\mathrm{res}}$. Therefore Eq.~(\ref{taunorm}) may be rewritten in the following way
\begin{eqnarray}\label{taugen}
\tau=\tau_0+\tau_{\rm ret}\,,
\end{eqnarray}
where
\begin{eqnarray}\label{tauinst}
\tau_0=\frac{\langle \min (t_{\mathrm{hit}},t_{\mathrm{res}}) \rangle}{1-P_i}
\end{eqnarray}
is the MFPT for instantaneous resetting, and
\begin{equation}\label{tauret}\tau_{\rm ret} = \frac{\langle t_{\mathrm{ret,b}} \rangle P_i}{1-P_i}. \end{equation} 
is additional term, accounting for the return motion. Compared to Eq.~(2) of Ref.~\cite{shlomi}, Eq.~(\ref{taugen}) does not contain the third term, corresponding to staying home. 
Eqs.~(\ref{tauinst}-\ref{tauret}) are, however, awkward for the further usage. Let us perform some modifications of these expressions.

Let $p(x|t;b)$ be the PDF of the particle's positions on a semi-infinite interval with an absorbing boundary (target) at $b > 0$. Then $\Omega(t) = \int_{-\infty}^{b} p(x'|t;b)$ is the survival probability of the particle 
in the interval up to time $t$, and $\Omega(t_{\mathrm{res}})$ the probability that no hitting took place before resetting at time 
$t_{\mathrm{res}}$. The hitting time density now can be expressed according to $\omega(t) = - \frac{d}{dt} \Omega(t)$. In \cite{SokChech} it has been shown that
\begin{equation}\label{mintt}
\langle \min (t_{\mathrm{hit}},t_{\mathrm{res}}) \rangle=\int_0^\infty dt \Psi(t) \Omega(t)\,. 
\end{equation}
The probability of an idle run is 
\begin{equation}\label{pipi}
P_i=\int_0^\infty dt \psi(t) \int_{-\infty}^{b} p(x|t;b) dx\,.
\end{equation}
Using Eq.~(\ref{mintt}-\ref{pipi}) we get for the MFPT for the instantaneous resetting \cite{SokChech}:
\begin{equation}\label{tau0wow}
\tau_0 = \frac{ \int_0^\infty dt \Psi(t) \Omega(t)}{1-\int_0^\infty dt \psi(t) \int_{-\infty}^{b} p(x|t;b) dx }\,,
\end{equation}
It is easy to see that
\begin{equation}\label{ttret}
\langle t_{\mathrm{ret,b}} \rangle P_i=\int_0^\infty  \psi(t)dt \int_{-\infty}^{b} t_{\mathrm{ret}}
p(x|t;b) dx .
\end{equation}
Thus, the additional term, accounting for the return motion, is
\begin{equation}\label{tauwowret}
\tau_{\rm ret} = \frac{\int_0^\infty  \psi(t)dt \int_{-\infty}^{b} t_{\mathrm{ret}}
p(x|t;b) dx}{1-\int_0^\infty dt \psi(t) \int_{-\infty}^{b} p(x|t;b) dx }\,.
\end{equation}
The return time $t_{\mathrm{ret}}$ is given by Eqs.~(\ref{ttretv}),~(\ref{ttreta}) and~(\ref{tretspring}) for return at constant velocity, constant acceleration and under the action of the harmonic force, correspondingly.

For the case of the Brownian motion our $p(x|t;b)$ reads
\begin{equation}
p(x|t;b) = \frac{1}{\sqrt{4 \pi D t}} \left[\exp \left(-\frac{x^2}{4 D t} \right) - \exp
\left(-\frac{(x-2b)^2}{4 D t} \right) \right]
\end{equation}
for $x < b$ and vanishes otherwise. The survival probability is 
\begin{equation}
 \Omega (t)=\mathrm{erf}\left(\frac{b}{2\sqrt{Dt}}\right)
\end{equation}
where $\mathrm{erf}(b)$ is the error function.

For exponential resetting the first term $\tau_0$  (Eq.~\ref{tau0wow}) represents the very well known expression for diffusion with instantaneous resetting \cite{EvansMajumdar}:
\begin{equation}\label{tau0exp}
\tau_0=(e^{\zeta}-1)/r.
\end{equation}
Here we have introduced a dimensionless parameter $\zeta=b\sqrt{r/D}$, accounting for the ratio between two main length parameters in the system:
the distance to the target and the characteristic diffusion length between two subsequent resetting events. The probability, that the target has been found during the run, is $P_i=1-e^{-\zeta}$.

The deterministic resetting has been shown to be the most effective search process \cite{SokChech, shlomi2017, palrt}.
In order to describe the MFPT, it is useful to introduce a dimensionless parameter $\xi=b/\sqrt{4Dt_r}$. In this case the MFPT for instantaneous resetting (Eq.~\ref{tau0wow}) takes the form
\begin{equation}\label{tau0det}
\tau_0=\frac{t_r\left(1+2\xi^2\right)
\mathrm{erf}\left(\xi\right)-2\xi^2t_r+\frac{2\xi t_r}{\sqrt{\pi}}\exp\left(-\xi^2\right)}{\mathrm{erfc}\left(\xi\right)}\,.
\end{equation}
The probability, that the target has been found during the run $P_i=\rm erf(\xi)$.

\begin{figure*}[ht]
\centerline{\includegraphics[width=1.\columnwidth]{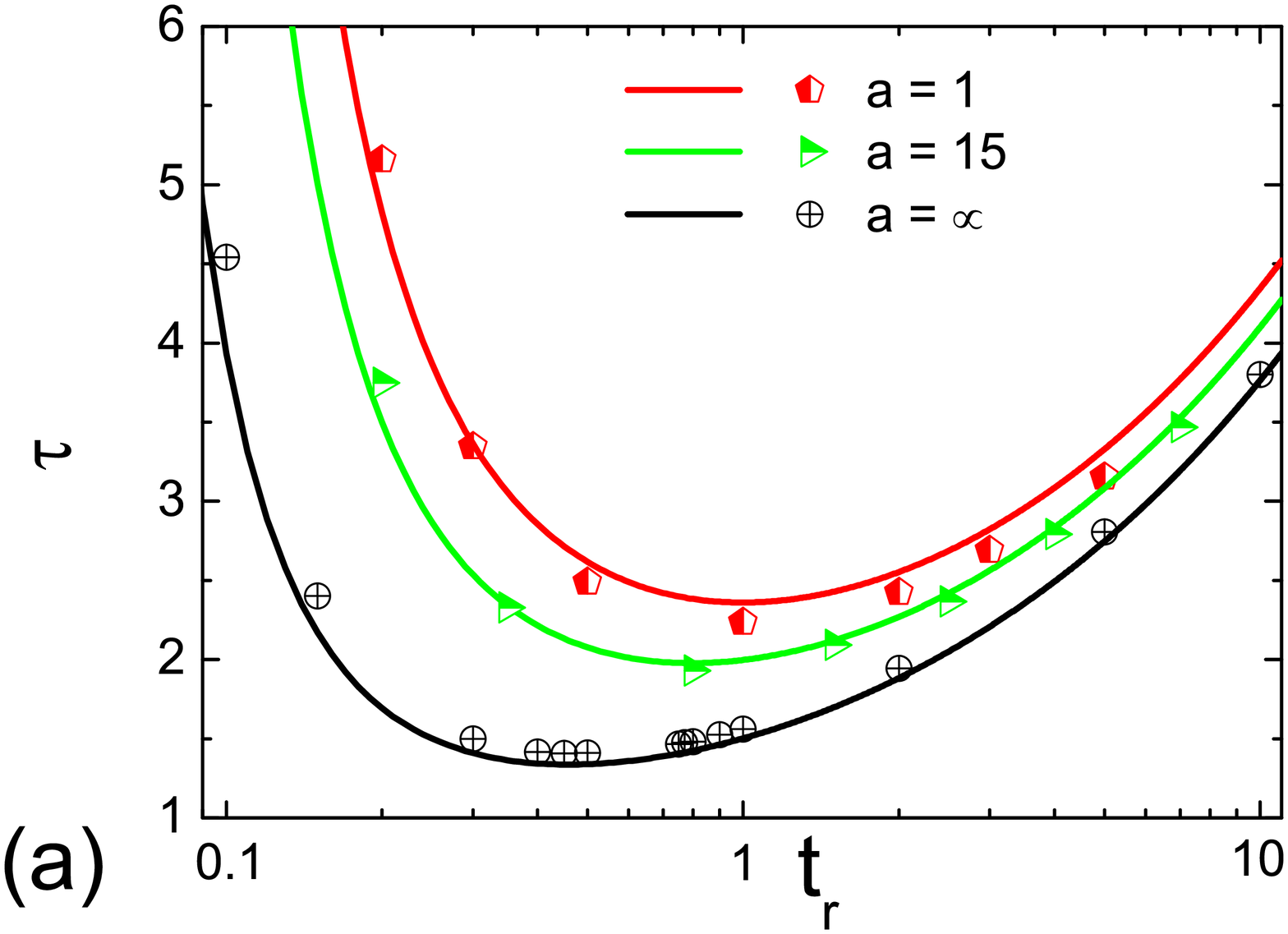}\includegraphics[width=1.\columnwidth]{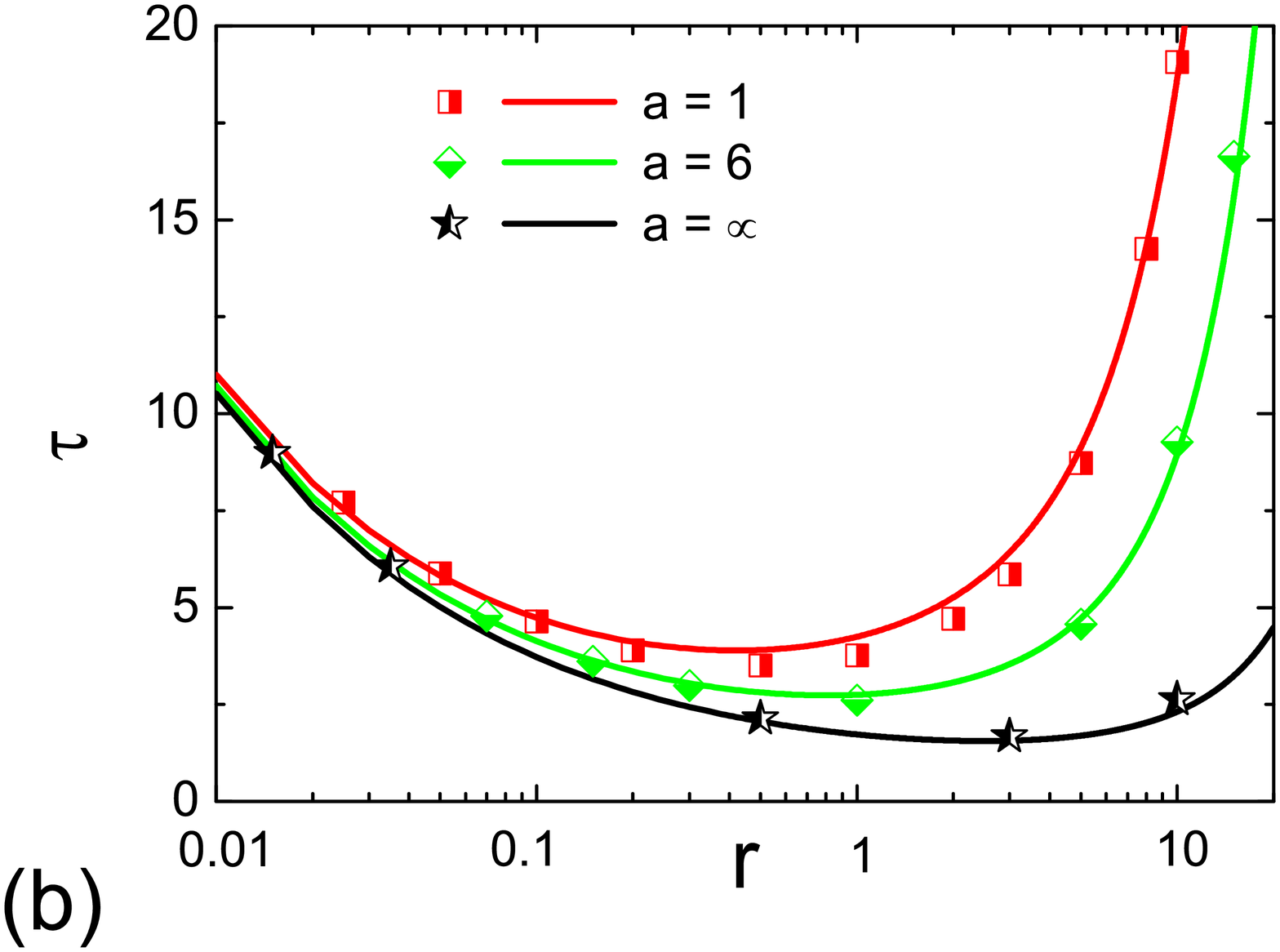}}
\caption{Mean first passage times for the return at constant acceleration for (a) deterministic resetting, $\psi(t)=\delta(t-t_r)$, and (b) exponential resetting, $\psi(t)=re^{-rt}$. Symbols correspond to the simulation data, solid lines -- to
the analytical result, Eq.~(\ref{taugen}) with $\tau_0$ and $\tau_{\rm ret}$ given by Eq.~(\ref{tau0det}) and Eq.~(\ref{ladet}) for the resetting at constant pace and Eq.~(\ref{tau0exp}) and Eq.~(\ref{laexp}) for the exponential resetting. The target is located at $b=1$.  $a=\infty$ corresponds to the case of instantaneous return.}
\label{Gfpta}
\end{figure*} 

\subsection{Return at a constant speed}

In the case of return with a constant velocity the full return time has the following form
\begin{equation}
\label{tauall1}
\tau=\tau_0+\frac{l_v}{v}.
\end{equation}
Here $l_v$ is the characteristic length of the return paths:
\begin{equation}\label{l0}
l_v = \frac{\int_0^\infty dt \psi(t) \int_{-\infty}^{b}| x |p(x|t;b) dx}{1-\int_0^\infty dt \psi(t) \int_{-\infty}^{b} p(x|t;b) dx }.
\end{equation}
In the case of exponential resetting the parameter $l_v$ is
\begin{equation}\label{l0exp}
l_v=b\left[\frac{1}{\zeta}\left( e^{\zeta}-e^{-\zeta}\right)-1\right]\,. 
\end{equation}
While the PDF for ordinary diffusion with exponential resetting does not depend on the return velocity, the MFPT has an explicit velocity-dependence. 
The MFPT for exponential resetting has been obtained independently in \cite{shlomi}. 

For the deterministic resetting the characteristic length is
\begin{equation}\label{l0det}
l_v= \frac{\frac{b}{\sqrt{\pi}\xi}\left(1-\exp\left(-4\xi^2\right)\right)+b\left(1+\mathrm{erf}\left(\xi\right)-2 
\mathrm{erf}\left(2\xi\right)\right)}{\mathrm{erfc}\left(\xi\right)} .
\end{equation}

The MFPT $\tau$ for both instantaneous return and for return at a constant speed are shown at Fig.~\ref{Gfpt}, demonstrating
good agreement between simulation results and theoretical predictions. The MFPT increases with
decreasing of the velocity of the ballistic motion towards the origin. The optimal resetting time $t_r$ for
the delta distribution of the resetting events increases with decreasing of $v$: the longer the duration of
the return period, the longer is the time period between subsequent resetting events. Similar effect has been
observed for the exponential distribution of the resetting events: the larger is the velocity $v$ the shorter
is the optimal average time between resetting events $1/r$, which corresponds to larger values of $r$.

\subsection{Return at a constant acceleration}

For the return at constant acceleration and the exponential resetting the correction due to the finite return time is equal to
\begin{eqnarray}\nonumber
&&\tau_{\rm ret}=\sqrt{\frac{\pi b}{2a\zeta}}\left[1-\frac{1}{\sqrt{\pi}}\Gamma\left(\frac32,\zeta\right)\right]e^{-\zeta}-\\
&&-\sqrt{\frac{\pi b}{8a\zeta}} e^{-3\zeta}\rm \left(1+erfi\left(\sqrt{\zeta}\right)\right)+\sqrt{\frac{b}{2a}}e^{-2\zeta}\,.
\label{laexp}
\end{eqnarray}
Here $\rm erfi(\zeta)$ is the imaginary error function. 

In the case of the deterministic resetting $\tau_{\rm ret}$ is given by the following expression
\begin{eqnarray}\label{ladet}
&&\tau_{\rm ret}=\left[\sqrt{\frac{b}{4\pi a\xi}}\left(2\Gamma\left(\frac34\right)-\Gamma\left(\frac34,\xi ^2\right)\right)-\right.\\\nonumber
&&-\frac{\xi\sqrt{b}}{\sqrt{2\pi a}}e^{- \frac{31\xi^2}{8}}\left(K_{\frac34}\left(\frac{\xi^2}{8}\right)-K_{\frac14}\left(\frac{\xi^2}{8}\right)\right)-\\
&&\left.-\frac{\sqrt{2b}\xi}{\sqrt{\pi a}}e^{-4\xi^2}\int_0^{1}dz\sqrt{z}\exp\left(-z^2\xi^2+z\xi^2\right)\right]/\rm erfc(\xi)\,.\nonumber
\end{eqnarray}
Here $K_a(\xi)$ is the modified Bessel function of second kind, and the last integral has to be evaluated numerically. 

The MFPT for the return at constant acceleration is depicted at Fig.~\ref{Gfpta}. The behavior of MFPT is similar to the case of the resetting at constant velocity. The MFPT grows with increasing of $r$ for exponential resetting and decreasing of $t_r$ for the deterministic resetting. The larger the acceleration the smaller the MFPT. For $a\to \infty$ the MFPT is the same as in the case of the instantaneous resetting.

\begin{figure*}[ht]
\centerline{\includegraphics[width=1.\columnwidth]{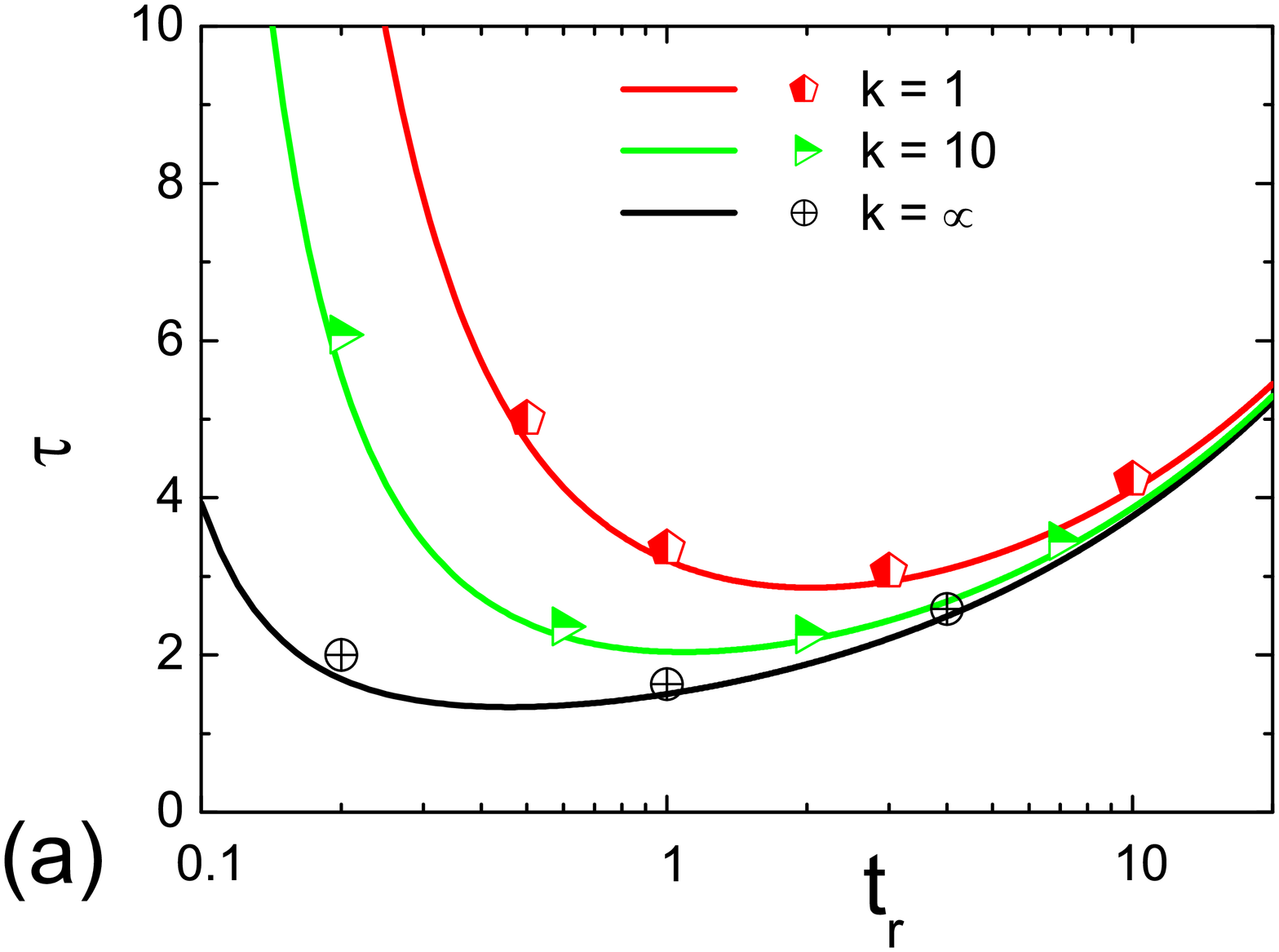}\includegraphics[width=1.\columnwidth]{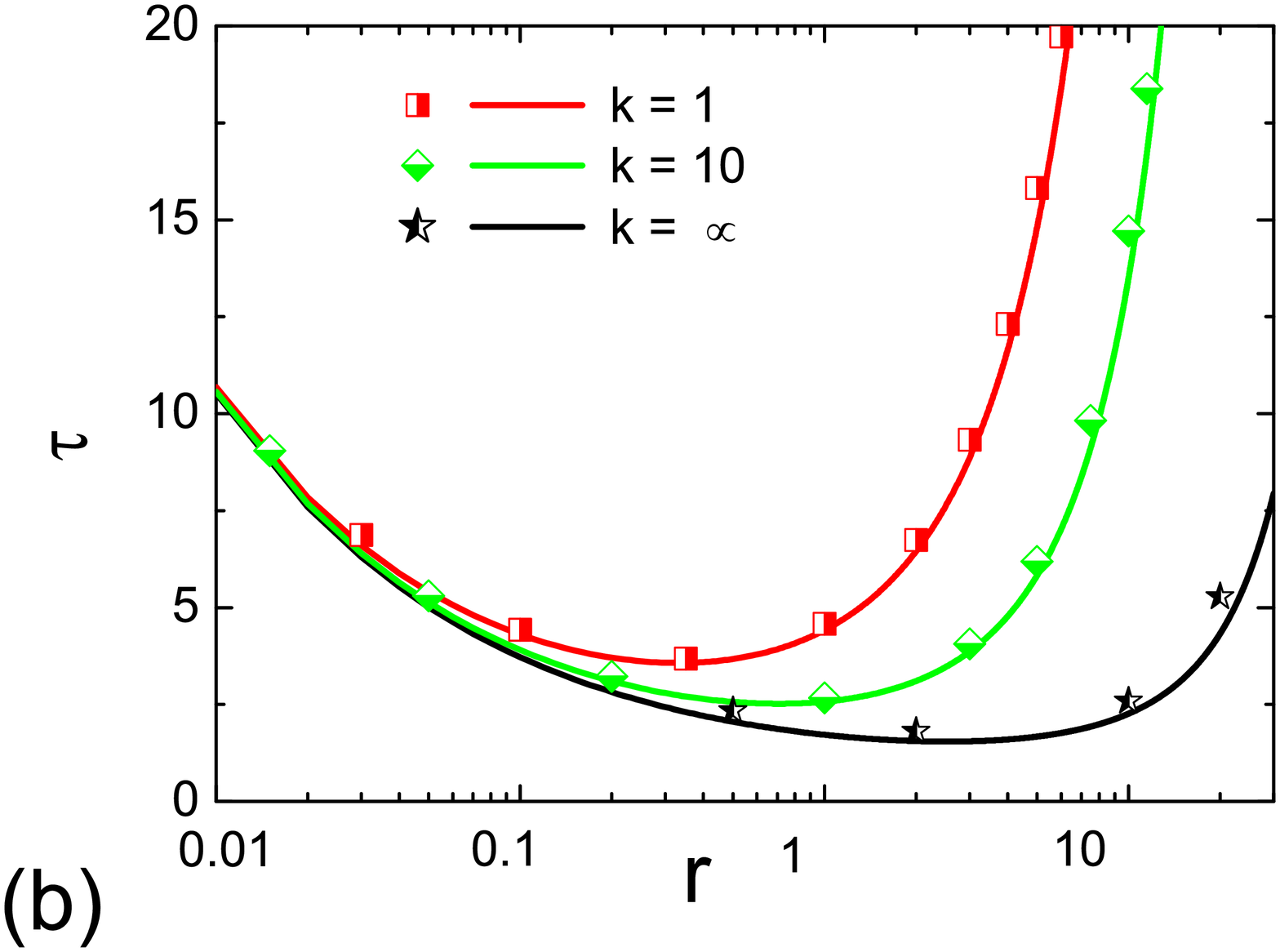}}
\caption{Mean first passage times for the return under the action of the harmonic spring . The particle moves with acceleration $a=-kx$ under (a) deterministic resetting, $\psi(t)=\delta(t-t_r)$, and (b) exponential resetting, $\psi(t)=re^{-rt}$. Symbols correspond to the simulation data, solid lines -- to
the analytical result, Eq.~(\ref{taugen}) with $\tau_0$ and $\tau_{\rm ret}$ given by Eq.~(\ref{tau0det}) and Eq.~(\ref{lkdet}) for the resetting at constant pace and Eq.~(\ref{tau0exp}) and Eq.~(\ref{lkexp}) for the exponential resetting. The target is located at $b=1$.  $k=\infty$ corresponds to the case of instantaneous return.}
\label{Gfptk}
\end{figure*} 

\subsection{Return under the action of a harmonic force}

Let us consider the MFPT for the return with the acceleration proportional to the distance between coordinate and the resetting point: $a=-kx$. The average return time is given by Eq.~(\ref{tretspring}) and does not depend on any parameters of resetting. The correction to MFPT for the exponential resetting it is equal to
\begin{equation}
\label{lkexp}
\tau_{\rm ret}=\frac{\pi}{2\sqrt{k}}\left(e^{\zeta}-1\right)\,.
\end{equation}
For the deterministic resetting
\begin{equation}
\label{lkdet}
\tau_{\rm ret}=\frac{\pi}{2\sqrt{k}}\frac{\rm erf\left(\xi\right)}{\rm erfc\left(\xi\right)} .
\end{equation}
The comparison between theory and simulations for MFPT is given at Fig.~\ref{Gfptk}. The smaller is the spring constant $k$, the larger is the MFPT.

\section{Conclusions.} 
We have considered a model of a resetting process, in which a particle, performing a random motion, returns
to its origin not immediately after the resetting event, but, instead, moves towards the origin following a given equation of motion.  Once the origin is reached, the particle starts performing the stochastic motion again.
In the present work we derived general expressions for the probability density function and mean squared displacement for this process. 

In the current study we have specifically discussed the case of the return at a constant speed, at a constant acceleration and under the action of a harmonic force. 
For a Brownian motion during the displacement phase, two specific resetting protocols were 
considered, for which analytical results can be obtained in a closed form: the exponential waiting time distribution 
between the beginning of the stochastic motion and the reset event, and a deterministic resetting after a fixed time since the beginning of the stochastic motion. Interestingly, for the Brownian motion with the exponential resetting, and for the return at a constant speed, at a constant acceleration, and, in general, with velocity depending only 
on the distance from the position the particle had at the end of the displacement phase, $v=v(x_0-x)$, the displacement's PDF does not depend on the corresponding velocity or acceleration. However, if the return velocity explicitly depends on $x$ and $x_0$ (and not only on their difference), this invariance does not hold anymore. The invariance also does not hold for the deterministic resetting.

We have addressed the question of the invariance of the PDF with respect to the return speed in a general setting.
While using our general expressions it is easy to prove whether a specific pair of displacement PDF in the transport phase
and waiting time PDF in the resetting process satisfies the condition for such an invariance, the problem of finding 
a waiting time PDF under which a stationary PDF under resetting will be invariant with respect to return speed for a given
$p(x|t)$ is tricky and worth further investigation. However, we discussed two simple examples showing that there exist situations in which such an invariance takes place under a whatever waiting time PDF, and also situations in which such an invariance can never hold. 

We have also considered the mean first passage times to a given target for Brownian motion under exponential and deterministic resetting. 
This mean first passage times depend on the return velocity, on the return acceleration or on the spring constant, correspondingly. The reason is that 
the first hitting cannot take place during the return phase which passes through a region which was already 
visited in the stochastic motion. Therefore, the smaller the velocity of return, the acceleration or the spring constant, the longer is the time necessary to 
find a target. As a consequence, the optimal time between the two resetting events increases with decreasing return speed, acceleration, or spring constant.

\appendix 

\section{Normalization of the probability distribution function}\label{Anorm}

The survival probability $\Phi(t)$ that the run didn't finish until time $t$ is
\begin{equation}
\Phi (t) = 1 - \int\limits_0^t {\phi (t')dt'}  = \int\limits_t^\infty  {\phi (t')dt'}\,.
\label{surv}
\end{equation}
Performing the integration of Eq.~(\ref{phirun}) over $t_{\rm run}$, one can get
\begin{equation}\label{eq:Phi}
 \!\! \!\! \Phi(t) =\int_0^\infty d t_{\mathrm{res}} \psi(t_{\mathrm{res}}) \int_{-\infty}^\infty d x_0 \Theta [t_{\mathrm{res}} 
+ t_{\mathrm{ret}}(x_0) - t] p(x_0|t_{\mathrm{res}}).  
\end{equation}

Let $t$ be the measurement time and $t_0$ the starting time of the last run. Then the probability density of $t_0$ is given by
\begin{equation}
  g(t_0) = \kappa(t_0) \Phi(t-t_0)\,.
 \label{eq:back}
\end{equation}
Here $\kappa(t_0)$ is the probability density that the run has started at time $t_0$ and $\Phi(t-t_0)$ accounts for the fact that it has not been terminated before $t$. The normalization of $g(t_0)$ (Eq.~\ref{eq:back}) follows by noting that in the Laplace domain $\tilde{\Phi}(s) = s^{-1}(1 - \tilde{\phi}(s))$. 
Therefore $I(t) = \int_0^t g(t_0) dt_0 = \int_0^t \kappa(t_0) \Phi(t-t_0) dt_0$ has a form of the convolution, 
and its Laplace transform $I(s) = \tilde{\kappa}(s) \tilde{\Phi}(s)$ is equal to $s^{-1}$, which corresponds to $I(t)=1$ 
in the time domain. Hence,
\begin{equation}
 \int_0^t g(t_0) dt_0 = \int_0^t \kappa(t_0) \Phi(t-t_0) dt_0 =1.
 \label{eq:Norm}
\end{equation}

Now from Eqs.~(\ref{qqq}-\ref{Q1Q2}) one can explicitly write down the equations for $Q_1(x;\Delta t)$ and $Q_2(x;\Delta t)$
\begin{eqnarray}\label{eq:conditioned}
&& Q(x;\Delta t) = Q_1(x;\Delta t)+Q_2(x;\Delta t)\,,\\
&&  Q_1 = \int_0^\infty p(x|\Delta t) \Theta(t_{\mathrm{res}} - \Delta t) \psi(t_{\mathrm{res}}) dt_{\mathrm{res}}\,,  \nonumber \\
&&  Q_2 = \int_0^\infty d t_{\mathrm{res}} \psi(t_{\mathrm{res}}) \int_{-\infty}^\infty dx_0 \delta[x - X(\Delta t-t_{\mathrm{res}};x_0)]\nonumber \\
&& \times\Theta(\Delta t-t_{\mathrm{res}} ) \Theta[t_{\mathrm{res}} + t_{\rm ret} (x_0) - \Delta t] p(x_0|t_{\mathrm{res}}). \nonumber
\end{eqnarray}

The fact that $P(x,t)$ (Eq.~\ref{eq:Poft}) is normalized can be seen by integrating Eq.~(\ref{eq:Norm}) over $x$. Integrating $Q(x; \Delta t)$, Eq.~(\ref{eq:conditioned}), over $x$ we get:
\begin{eqnarray}
&&\int_{-\infty}^\infty  Q(x;\Delta t) dx = \int_0^\infty  dt_{\mathrm{res}} \psi(t_{\mathrm{res}}) \left[ \Theta(t_{\mathrm{res}} - \Delta t) 
 + \right. \\ &&\left. + \Theta(\Delta t-t_{\mathrm{res}} )
 \int_{-\infty}^\infty dx_0  \Theta[t_{\mathrm{res}} + t_{\rm ret} (x_0) - \Delta t] p(x_0|t_{\mathrm{res}}) \right].\nonumber
\end{eqnarray}
Now we note that $\Theta(-y) = 1 - \Theta(y)$ for any real $y$. Making this substitution for the last $\Theta$-function in the last 
expression and using that $\int_{-\infty}^\infty dx_0 p(x_0|t_{\mathrm{res}}) = 1$ we get:
\begin{eqnarray}
&& \int_{-\infty}^\infty  Q(x;\Delta t) dx = \int_0^\infty  dt_{\mathrm{res}} \psi(t_{\mathrm{res}})\times\\\nonumber &&\times\left[ 1- \int_{-\infty}^\infty dx_0 \Theta[\Delta t-t_{\mathrm{res}} - t_{\rm ret} (x_0)]  p(x_0|t_{\mathrm{res}}) \right]= \\
 &&  \!\!= \!\!\int_0^\infty \!\! dt_{\mathrm{res}} \psi(t_{\mathrm{res}}) \int_{-\infty}^\infty dx_0 \Theta[t_{\mathrm{res}} + t_{\rm ret} (x_0) - \Delta t] p(x_0|t_{\mathrm{res}})\nonumber
\end{eqnarray}
which coincides with our expression for $\Phi(\Delta t)$, Eq.~(\ref{eq:Phi}). The overall normalization then follows from Eq.~(\ref{eq:Norm}).\\


\begin{thebibliography}{99}

\bibitem{review} M. R. Evans, S. N. Majumdar and G. Schehr, https://arxiv.org/abs/1910.07993.
\bibitem{chemistry} S. Reuveni, M. Urbach, and J. Klafter, Proc. Natl. Akad. Sci. U.S.A. 111, 4391 (2014).
\bibitem{bio1} T. Rotbart, S. Reuveni, M. Urbakh, Phys. Rev. E \textbf{92}, 060101 (2015).
\bibitem{bio2} T. Robin, S. Reuveni, M. Urbakh, Nat. Commun. \textbf{9}, 779 (2018).
\bibitem{biology} \'E. Rold\'an, A. Lisica, D. S\'anchez-Taltavull, and S.W. Grill, Phys. Rev. E \textbf{93}, 062411 (2016).
\bibitem{bio3} S. Budnar, K. B. Husain, G. A. Gomez, M. Naghibosidat, S. Verma, N. A. Hamilton, R. G. Morris, A. S. Yap, 
Developmental Cell, \textbf{49}, 894 (2019)
\bibitem{computerscience}  A. Montanari and R. Zecchina, Phys. Rev. Lett. \textbf{88}, 178701 (2002).
\bibitem{EvansMajumdar} M. R. Evans and S. N. Majumdar, Phys. Rev. Lett. \textbf{106}, 160601 (2011).
\bibitem{levy1} {\L}. Ku\'smierz, S. N. Majumdar, S. Sabhapandit, and G. Schehr, Phys. Rev. Lett. \textbf{113}, 220602 (2014).
\bibitem{levy2} {\L}. Ku\'smierz, E. Gudowska-Nowak. Phys. Rev. E \textbf{92}, 052127 (2015).
\bibitem{MV2013}  M. Montero and J. Villarroel, Phys. Rev. E \textbf{87}, 012116 (2013).
\bibitem{MC2016} V. M$\rm\acute{e}$ndez and D. Campos, Phys. Rev. E \textbf{93}, 022106 (2016).
\bibitem{Sh2017}  V.P. Shkilev, Phys. Rev. E \textbf{96}, 012126 (2017).
\bibitem{ctrw} M. Montero, A. Mas-Puigdell\'osas, J. Villarroel. Eur. Phys. J. B \textbf{90}, 176 (2017).
\bibitem{Anna01} A.S. Bodrova, A.V. Chechkin, I.M. Sokolov. Phys. Rev. E \textbf{100}, 012119 (2019).
\bibitem{Anna02} A.S. Bodrova, A.V. Chechkin, I.M. Sokolov. Phys. Rev. E \textbf{100}, 012120 (2019). 
\bibitem{shlomi2017} A. Pal and S. Reuveni. Phys. Rev. Lett. \textbf{118}, 030603 (2017).
\bibitem{palrt} A. Pal, A. Kundu and M. R. Evans, J. Phys. A: Math. Theor. \textbf{49}, 225001 (2016).
\bibitem{NagarGupta} A. Nagar and S. Gupta, Phys. Rev. E \textbf{93}, 060102 (2016).
\bibitem{res2016} S. Eule and J. J. Metzger, New J. Phys. \textbf{18}, 033006 (2016).
\bibitem{shlomi2016} S. Reuveni. Phys. Rev. Lett. \textbf{116}, 170601 (2016).
\bibitem{redner} S. Redner, A Guide to First-Passage processes. Cambridge University Press, Cambridge, UK (2007).
\bibitem{redner2} R. Metzler, G. Oshanin and S. Redner, eds., First Passage Phenomena and Their Applications. World Scientific, Singapore (2014).
\bibitem{EM2011} M. R. Evans and S. N. Majumdar, J. Phys. A: Math. Theor. \textbf{44}, 435001 (2011).
\bibitem{evma13} M.R. Evans, S.N. Majumdar and K. Mallick. J. Phys. A: Math. Theor. \textbf{46}, 185001 (2013).
\bibitem{high12} M.R. Evans and S.N. Majumdar. J. Phys. A: Math. Theor. \textbf{47}, 285001 (2014).  
\bibitem{bhat} U. Bhat, C. de Bacco, S. Redner. J. Stat Mech. 083401 (2016).
\bibitem{SokChech} A.V. Chechkin, I.M. Sokolov, Phys. Rev. Lett. \textbf{121}, 050601 (2018).
\bibitem{wait1} A. Maso-Puigdellosas, D. Campos and V. Mendez. Phys. Rev. E \textbf{99} 012141 (2019).
\bibitem{wait2} A. Maso-Puigdellosas, D. Campos and V. Mendez, J. Stat. Mech. 033201 (2019).
\bibitem{wait3} M.R. Evans and S.N. Majumdar, J. Phys. A \textbf{52}, 01LT01 (2019).
\bibitem{shlomi} A. Pal, {\L}. Ku\'smierz, S. Reuveni, https://arxiv.org/abs/1906.06987.
\bibitem{shlomi1} A. Pal, {\L}. Ku\'smierz, S. Reuveni, New J. Phys. \textbf{21}, 113024 (2019).
\bibitem{shlomi2} A. Pal, {\L}. Ku\'smierz, S. Reuveni, Phys. Rev. E \textbf{100}, 040101(R) (2019).
\bibitem{sokbook} J. Klafter and I.M. Sokolov, First Steps in Random Walks: From Tools to Applications. Oxford University Press, New York, USA (2011).

\end{thebibliography}
 \end{document}